\documentclass[prd,amsmath,notitlepage,twocolumn,10pt]{revtex4-2}
\usepackage{graphicx}
\usepackage{float}
\usepackage{epstopdf,cancel}
\usepackage{epsf,latexsym,bbm,euscript}
\usepackage{amssymb,amsmath}
\usepackage{mathtools} 
\usepackage{times,graphics}
\usepackage{soul,xcolor}
\usepackage{mathtools}
\usepackage{braket}
\usepackage[normalem]{ulem}
\usepackage{setspace}
\usepackage{url,hyperref}
\hypersetup{colorlinks,linkcolor={blue!55!black},citecolor={red!45!black},urlcolor={blue!45!black},breaklinks=true}

\def\6{{\langle}}
\def\9{{\rangle}}
\def\cO{\mathcal{O}}

\newcommand\ContFracOp{%
	\operatornamewithlimits{%
		\mathchoice
		{\vcenter{\hbox{\huge $\mathbf{K}$}}}
		{\vcenter{\hbox{ $\mathbf{K}$}}}
		{\mathbf{K}}
		{\mathbf{K}}}}

\usepackage{mathtools}
\newcommand{\defeq}{{\vcentcolon=}}

\newcommand{\be}{\begin{equation}}
\newcommand{\ee}{\end{equation}}
\newcommand{\p}{\partial}

\usepackage{scalerel}
\usepackage{tikz}
\usetikzlibrary{svg.path}
\definecolor{orcidlogocol}{HTML}{A6CE39}
\tikzset{
	orcidlogo/.pic={
		\fill[orcidlogocol] svg{M256,128c0,70.7-57.3,128-128,128C57.3,256,0,198.7,0,128C0,57.3,57.3,0,128,0C198.7,0,256,57.3,256,128z};
		\fill[white] svg{M86.3,186.2H70.9V79.1h15.4v48.4V186.2z}
		svg{M108.9,79.1h41.6c39.6,0,57,28.3,57,53.6c0,27.5-21.5,53.6-56.8,53.6h-41.8V79.1z M124.3,172.4h24.5c34.9,0,42.9-26.5,42.9-39.7c0-21.5-13.7-39.7-43.7-39.7h-23.7V172.4z}
		svg{M88.7,56.8c0,5.5-4.5,10.1-10.1,10.1c-5.6,0-10.1-4.6-10.1-10.1c0-5.6,4.5-10.1,10.1-10.1C84.2,46.7,88.7,51.3,88.7,56.8z};
	}
}
\newcommand\orcidlink[1]{\href{https://orcid.org/#1}{\mbox{\scalerel*{
				\begin{tikzpicture}[yscale=-1,transform shape]
				\pic{orcidlogo};
				\end{tikzpicture}}{X}}}}

\begin{document}

\title{Semi-classical imprints on quasinormal mode spectra}

\author{Fil Simovic\,\orcidlink{0000-0003-1736-8779}}
\email{fil.simovic@mq.edu.au}

\author{Daniel R. Terno\,\orcidlink{0000-0002-0779-0100}}
\email{daniel.terno@mq.edu.au}

\affiliation{School of Mathematical and Physical Sciences, Macquarie University, NSW 2109, Australia}

\begin{abstract}
	\vspace*{1mm}

	We compute quasinormal mode frequencies for static limits of physical black holes---semi-classical black hole solutions to Einstein-Hilbert gravity characterized by the finite formation time of an apparent horizon and its weak regularity. These assumptions lead to a highly constrained yet non-trivial form of the metric and components of the energy-momentum tensor near the horizon, which contain as a special case many known models of black holes. Using a two-point M-fraction approximation to construct an interpolating metric which captures the essential near-horizon and asymptotic properties of black holes, we explore a large part of the parameter space that characterizes the near-horizon geometry. We cast  the perturbation problem as a discretized homogeneous eigensystem and compute the low-lying quasinormal mode frequencies for perturbations of a massless scalar field. Working in spherical symmetry, we provide rough constraints on leading and subleading deviations from the Schwarzschild solution which arise in a semi-classical setting, and demonstrate the potential for mimicking signatures of axially-symmetric geometries from a spherically-symmetric source.
	
	\medskip

\end{abstract}

\maketitle

\section{Introduction}

Gravitational waves (GWs) are a highly effective tool for probing the structure of massive compact objects, especially black holes \cite{LIGO:21}. GWs detected on Earth are emitted during violent processes involving  massive compact objects. As they settle into a steady state, one component of the GW signal is largely determined by the quasinormal modes (QNMs). These correspond to a relaxation to equilibrium in the perturbative regime after an astrophysical object is disturbed \cite{RW:57,P:71,C:92,FN:98}. QNMs provide a sensitive signature of the physical properties of compact objects and can reveal details of the underlying physical theory if it differs from general relativity \cite{BCNS:19}. For the uncharged Kerr black hole, this spectral fingerprint is uniquely determined by the total mass $M$ and angular momentum $J$, thus allowing for tests of the no-hair theorem, alternative theories of gravity, as well as providing an independent channel to estimate the mass and spin of a black hole, complementing other sources \cite{RN:03,B:17,EHT:19,CP:19,IM:19}.

Determined entirely by the black hole's structure through its defining charges and the spacetime's asymptotics, quasinormal modes offer a way to test different black hole models. The Schwarzschild and Kerr solutions, though they provide the simplest models consistent with current observations of astrophysical black hole candidates, face conceptual difficulties inherent in their construction—chief among them being the teleological nature of the event horizon and the presence of a singularity behind it \cite{HE:73,BCNS:19,CP:19,MMT:22}.  These difficulties are absent by design in certain models of ultra-compact objects (UCOs), including regular black holes and horizonless UCOs. From this point of view, black holes are part of a hierarchy of  UCOs  characterized by their compactness and the presence of a photon sphere \cite{BCNS:19,CP:19}.

Once gravitational collapse proceeds beyond the density of a neutron star, from the perspective of a distant observer  there are three logical possibilities: (a) perpetual ongoing collapse, with the compactness parameter approaching zero asymptotically ($t \to \infty$),  thus a horizon being an asymptotic concept; (b) the formation of a transient or stable object, where the compactness parameter reaches a minimal non-zero value, possibly asymptotically; (c) the formation of a light-trapping region  in finite time according to the clock of a distant observer \cite{MMT:22}.

 At the same time, the dynamics of the final stages of collapse remains poorly understood. On the one hand, quantum effects are routinely handled within the framework of semiclassical gravity. It is expected to break down in the presence of large curvature. On the other hand, quantum effects may play a significant role even at the horizon scale of macroscopic black holes \cite{A:20} or prevent horizon formation altogether \cite{C-R:18,BBC:21}.

With the aim of both understanding potential observational signatures of such effects, and providing a useful framework for testing various models of black hole and ultracompact objects, this work takes some initial steps towards understanding how semi-classical effects near the horizon imprint themselves on the QNM spectrum. This will serve as a precursor for studying quasinormal modes of fully dynamical semi-classical black holes whose
near-horizon geometry   is constrained by the requirements of regularity and finite formation time \cite{MMT:22,BMMT:19,DSST:23}.

To do so, we employ a two-point Padé approximation to interpolate between a semi-classically sourced near-horizon black hole geometry and the desired asymptotic conditions. Combined with a matrix discretization method, such a scheme provides a useful way to study QNMs in a variety of settings. We determine how leading and sub-leading deviations from the Schwarzschild geometry imprint themselves on the QNM spectrum, and provide some preliminary constraints on the components of the renormalized energy-momentum tensor near the horizon using GW data from LIGO-VIRGO observations of binary black hole coalescence.

Our procedure differs from previous work in this area in several important ways. First, semi-classical treatments often rely on simplifying approximations to the energy-momentum tensor (EMT), which has not been computed for generic fields in four dimensions. We work in a setting where no assumptions are made about the quantum state and all matter content is treated jointly, allowing us to provide general constraints at the expense of a precise characterization of the EMT and geometry. Other studies examining QNMs in quantum settings have focused on specific metrics which are tied to underlying assumptions about the correct quantization of gravity \cite{L:20,G:24,J:23}, limiting their applicability to those settings. And while Padé-like approximants have been employed previously in the study of quasinormal modes, namely in \cite{RZ:14,KZ:22}, their approximation was designed to incorporate parameterized post-Newtonian (PPN) corrections, while we seek to isolate the contribution of near-horizon effects to the spectrum and present a more constrained form of the metric. To facilitate the interpolation we use a related, but distinct, M-fraction family of Pad\'{e} approximants. Even in its  preliminary form our scheme shows that, at least when dealing with restricted observational information, the QNMs of more general spherically-symmetric metrics have the potential to mimic the expected signatures of Kerr black holes. Finally, though stated in the context of semi-classical gravity, the method of analyzing QNMs presented here can be easily modified and applied to a wide variety of spacetimes in a purely classical setting.

This paper is organized as follows. In Section \ref{qnmintro} we present the necessary background to discuss quasinormal modes. In Section \ref{k1} we describe the near-horizon metrics we consider. In Section \ref{Padé}, we discuss the boundary conditions,  construct the required interpolating function and derive the radial wave equation. In Section \ref{matrix} we present the discretization scheme used to numerically compute the quasinormal frequencies. In Section \ref{results} we compute the quasinormal frequencies of scalar perturbations for the low-$l$ fundamental modes, and provide estimates on the bounds of the deviation parameters from GW observations. We conclude with some comments on limitations and directions for future work. Natural units where $G=c=\hbar=1$ are used throughout.

\section{Quasinormal modes}\label{qnmintro}

The study of QNMs is extensively covered in the literature. Excellent reviews can be found in \cite{BCS:09,N:99,KZ:11,C:92,FN:98}, and here we provide only a brief summary of the relevant facts.

Assuming the dynamics of interest occur in the linear regime, small perturbations around the   metric $g_{ab}^0$ and matter fields $\Phi^0$  are given by
\be\nonumber
g_{\mu\nu}=g_{\mu\nu}^0+\delta g_{\mu\nu}\ ,\quad \Phi=\Phi^0+\delta\Phi\ ,
\ee
where $|\delta g_{\mu\nu}|\ll |g_{\mu\nu}^0|$ and $\|\delta\Phi\|\ll \|\Phi^0\|$.
Five types of perturbations generically arise: those of  scalar, Dirac, and electromagnetic fields, as well as scalar (Zerilli) and vector (Regge--Wheeler) gravitational perturbations.

We consider the simplest possible case of a massless scalar field $\Phi$ on  a time-independent spherically-symmetric background $g_{\mu\nu}\equiv g^0_{\mu\nu}$, working in the frequency domain. Its perturbations are a natural first-pass context in which to study new metrics, providing a simplified model where many of the expected features of higher-spin spectra can be studied \cite{FN:98,KZ:11,RZ:14,BCCD:22,FLV:24,K:23}. Moreover, scalar fields have previously been implicated as dark matter candidates, and are known to allow for long-lived configurations near an astrophysical black hole, providing a possible manifestation of dark matter halos \cite{B:11,H:12}. For simplicity  we also restrict ourselves to standard general relativity (GR) defined by the Einstein--Hilbert action. The field equation for $\Phi$ is thus
\begin{align}\label{pot}
\square\Phi=&\frac{1}{\sqrt{-g}} \partial_\mu\left(g^{\mu \nu} \sqrt{-g} \partial_\nu \Phi\right)=0\ ,
\end{align}
and perturbations are separable in the form
\be\nonumber
\Phi(t,r,\theta,\phi)=\sum_{lmn}\dfrac{1}{r}\,\phi_{lmn}(r) Y_{lmn}(\theta,\phi)e^{-i\omega_{lmn} t}\ ,
\ee
where $m$ is the azimuthal number, $n$ is the overtone number, and $Y_{lmn}$ are   the spherical harmonics. This ansatz reduces the Klein-Gordon equation to an angular equation and a Schr\"{o}dinger-like Regge--Wheeler  equation for the radial part of the field,
\be\label{kg1}
\dfrac{\p^2\phi}{\p r_*^2}+(\omega^2-V)\phi=0\ ,\quad V=f\left(\dfrac{l(l+1)}{r^2}+\dfrac{f'}{r}\right)\ ,
\ee
where $f$ is the metric function (to be defined), $V$ is the effective potential, and primes denote derivatives with respect to $r$ throughout. The tortoise coordinate $r_*$ is  defined by
\be
dr_*=  {dr}/{f(r)}\ .
\ee
This expression is not analytically integrable in the general case, although it is for the backgrounds we consider here. Nevertheless, it will still be convenient to work directly with the radial coordinate $r$, in terms of which the radial equation \eqref{kg1} is
\be\label{radial1}
f\,\dfrac{\p^2\phi}{\p r^2}+ff'\dfrac{\p\phi}{\p r}+(\omega^2-V)\phi=0\ .
\ee
In the presence of a horizon, the oscillatory solutions are a subset of solutions of Eq.~\eqref{radial1} satisfying a purely ingoing boundary condition at the horizon,
\be\label{bcin}
\phi\sim e^{-i\omega r_*}\quad \text{as}\quad r_*\rightarrow -\infty\ ,
\ee
and a purely outgoing boundary condition in the asymptotic region,
\be\label{bcout}
\phi\sim e^{i\omega r_*}\quad \text{as}\quad r_*\rightarrow +\infty\ .
\ee

Enforcing these boundary conditions selects solutions that correspond to damped quasinormal oscillations arising from an initial perturbation (hence the name QNM). The frequency of these oscillations is independent of the initial amplitude. They are characterized by the physical frequency $f$ and damping time $\tau$ of each damped sinusoid present in the signal, and are related to the complex frequency $\omega$ through
\be\nonumber
\omega_{lmn}=2\pi f_{lmn}+\dfrac{i}{\tau_{lmn}}\ .
\ee

For fixed $a\defeq J/M$ the complex frequency scales as $\omega\sim M^{-1}$, so it is convenient to define a dimensionless frequency
\be
\Omega_{lmn}=M\omega_{lmn}\ .\nonumber
\ee
For an unknown mass $M$  the result of observation of a frequency $\Omega$ is given by a straight line in the $\text{Re}(\Omega)-\text{Im}(\Omega)$ plane parameterized by $M$, which will intersect various  $\Omega_{lm}$ curves which are themselves parameterized by $a$. Thus, knowing only $f$ and $\tau$ one cannot uniquely identify the black hole mass and angular momentum $J$, but can only reduce the possibilities to a (possibly countably infinite) set of $(M,a)$ pairs. If one knows they have observed a specific $(lmn)$ mode, then $M$ and $a$ are determined by the unique intersection.

Deviations from the expected frequency and damping time based on the vacuum Kerr solution of general relativity can be conveniently parameterized in a theory-agnostic way using the scheme developed in \cite{L:12,GVS:12,G:21}. The model is constructed assuming quasinormal frequencies depend on additional independent dimensionless parameters or charges $\{\Delta \hat{\omega}_{lm}\}$ along with $(M,J)$, which may arise from additional matter/gauge field couplings or corrections to the Einstein--Hilbert action. The observed frequency and damping time can then be represented as fractional deviations from their predicted Kerr values:
\begin{align}\nonumber
	& f_{\ell m 0}=f_{\ell m 0}^{\mathrm{GR}}(1+\delta \hat{f}_{\ell m 0}), \\
	& \tau_{\ell m 0}=\tau_{\ell m 0}^{\mathrm{GR}}\left(1+\delta \hat{\tau}_{\ell m 0}\right) \nonumber
\end{align}
As described in \cite{A:22}, a detailed study of high-SNR black hole binary coalescence signals allows one to constrain the fractional deviations from the Kerr value of the $(220)$ mode to be
\be\label{constraint}
\delta \hat{f}_{220}=0.02_{\,-0.03}^{\,+0.03} ,\quad \delta \hat{\tau}_{220}=0.13_{\,-0.11}^{\,+0.11}\ .
\ee
where the bounds encompass the 90\% confidence interval. We will be interested in relating deviations in quasinormal frequencies of a similar order of magnitude to changes in the underlying metric parameters near the horizon,
using the scalar perturbations as a stand-in for more complicated gravitational perturbation. It is the behaviour of the latter that is related to the observed GW signals.

\section{Near-horizon metrics}\label{k1}

In this work, the QNM analysis relies on several key assumptions of semi-classical gravity which we assume to be valid throughout the paper. First, regardless of the role that is played by quantum effects,  classical geometric notions such as a metric and curvature, as well as various notions of horizons, are well-defined. Second,  the motion of test particles and fields on such a background geometry is determined by the relevant classical equations of motion.

Semi-classical general relativity is then completed by taking the renormalized expectation value of the EMT as the source for Einstein's equations:
\be
G_{\mu\nu}=8\pi\bra\psi\hat{T}_{\mu\nu}(\hat\Phi,\hat\Pi,g)\ket\psi_{ \text{ren}}\equiv 8\pi T_{\mu\nu}
\ee
Here $G_{\mu\nu} \defeq R_{\mu\nu}-\tfrac{1}{2} Rg_{\mu\nu}$ is the Einstein tensor, and the EMT in general depends on all quantum fields $\hat\Phi$ and their conjugate momenta $\hat\Pi$ as well on the classical metric $g_{\mu\nu}$. The evolution of the field state $|\psi\9$ is supposed to be governed by the Schr\"{o}dinger equation with the Hamiltonian $\hat H[\hat\Phi,\hat\Pi;g]$.  The EMT may contain  curvature  counterterms that arise from renormalisation, as well as the cosmological constant, or higher-order curvature terms  of modified metric theories of gravity.  Though the quantum contribution to the right-hand side is locally suppressed by a factor of $\hbar$ relative to the classical part of $T_{\mu\nu} $,  there are a variety of settings in which these corrections can become large and  affect both the geometry of a black hole and its formation dynamics.

One of our goals is to map the space of parameters which define the near-horizon geometry to the frequencies of QNMs, with the goal of being maximally model-independent and sensitive to the largest possible domain for each of the defining parameters. The methods we describe also represent a first step towards extracting the observational signatures of the metrics that are obtained within the self-consistent approach to black hole horizons \cite{MMT:22}. This approach aims to obtain the  information about the near-horizon  region  and the EMT in it based on a minimal set of assumptions. Explicitly, we assume the validity of semi-classical physics as defined above, and supplement it by the requirements of horizon observability and weak regularity. The former is expressed as a requirement of  formation of trapped regions of spacetime in finite time according to distant observers.  The weak regularity demands only that   curvature scalars that are built by using   algebraic operations on the Riemann tensor components are finite on  their boundary. On the other hand, no assumptions about the EMT, large scale structure of the spacetime, or the underlying quantum states are necessary. Under these conditions, the restriction of spherical symmetry allows for an exhaustive classification of the allowed near-horizon geometries \cite{MMT:22}. We present the summary of the relevant facts and classification of admissible geometries in Appendix \ref{appA}.

A general spherically-symmetric metric  in Schwarzschild coordinates is given by
\be
ds^2=-e^{2h(t,r)}f(t,r)dt^2+f(t,r)^{-1}dr^2+r^2d\Omega_2\ , \label{staticmet}
\ee
where $r$ is the areal radius and $f=1-2M(t,r)/r$ is a scalar function with $M$ being the Misner--Sharp--Hernandez (MSH) mass. If the spacetime contains a black hole, its outer boundary --- the outer apparent horizon --- is given by the  largest (on a non-cosmological scale) real root of $f=0$, in all foliations that respect spherical symmetry. Admissible solutions are classified according to  the scaling of the effective EMT components, such as $\tau_t\defeq e^{-2h}T_{tt}$ (Eqs.~\eqref{split1}--\eqref{split3}), where $\tau_t\propto f^k$   on the approach to the horizon. Static limits of the dynamic solutions belong to the class $k=1$.
Well-known solutions, such as the Reissner--Nordstr\"{o}m, Bardeen, Hayward, and Simpson--Visser black holes belong to this class. Many popular models (including these) assume $h\equiv 0$, and to simplify the exposition in the following we only consider such solutions.

A near-horizon expansion in terms of ${\tilde x}\defeq r-r_g$ is constrained to have the form
\be\label{met}
f(\tilde x)\sim \alpha_1\frac{\tilde x}{r_g} +\alpha_2\frac{\tilde x^2}{r_g^2}+\mathcal{O}(\tilde x^{5/2}) \ ,
\ee
where the a priori absence of higher half-integers powers can be guaranteed only if additional regularity requirements  are imposed. In practice, all known  solutions contain only integer powers of $\tilde x$. The leading coefficient is related to the energy density at the horizon through
\be
\alpha_1=1-8\pi E\, r_g^2\ ,\label{a}\\
\ee
where $E=\rho(r_g)$, $E\leqslant1/8\pi r_g^2$, and higher order terms depend on the higher order terms in the EMT (see Appendix \ref{appA}). The vacuum  Schwarzschild metric corresponds to the values
\be\label{schwlim}
\alpha_1=1\ ,\quad a_2=-1\ ,
\ee
in \eqref{met}. As another example, consider the Bardeen metric
\be
f(r)=1-\frac{2M r^2}{(r^2+l^2)^{3/2}}\ ,
\ee
where $M$ is the Arnowitt–Deser–Misner (ADM) mass, and $l\defeq\lambda M$ is the regularisation parameter. Then for $\lambda\ll 1$ the horizon is located at
\be
r_g=2M\left( 1-\tfrac{3}{8}\lambda^2\right) +\cO\big(\lambda^4\big)\ ,
\ee
and
\be
\alpha_1=1-\frac{3}{4}\lambda^2+\cO\big(\lambda^4\big)\ , \quad \alpha_2=-1+ \frac{15}{8}\lambda^2+\cO\big(\lambda^4\big)\ .
\ee

It is easy to check which energy conditions \cite{HE:73} are satisfied for metrics of this type. The null energy condition at the horizon is violated for $\alpha_2<-1$, and the weak energy condition is violated there if $E<0$, i. e. $\alpha_1>1$.

\section{Boundary conditions and Pad\'{e} approximants} \label{Padé}

The boundary conditions of Eqs.~\eqref{bcin} and \eqref{bcout} only require the asymptotic form of the metric. Hence its behaviour of the metric near the horizon that is given by Eq.~\eqref{met} and in the asymptotic region are sufficient to specify an ansatz to the wave equation \eqref{radial1}. For all spherically-symmetric metrics the leading term in the far field expansion $(r\rightarrow \infty$) is
\be\label{far}
f(r)=1-\frac{2M}{r}+\cO(r^{-2}),
\ee
where $M$ is the ADM mass. It is convenient to express the difference between the gravitational radius and the ADM mass as \cite{RZ:14}
\be
\epsilon\defeq\frac{2M}{r_g}-1\ ,
\ee
which is (loosely) lower-bounded by neutron star compactness to be $\epsilon \gtrsim -0.4$ \cite{KZ:22}. Both Schwarzschild and Bardeen black holes have $\epsilon=0$. 

The terms of order $r^{-2}$  originate from deviations of the underlying theory from GR or from the presence of electric/magnetic charge. The first sub-leading term in the Bardeen model is of order $r^{-3}$, and in the Hayward model is of order $r^{-4}$.

To obtain the effective potential $V$ it is necessary to define the metric over the entire radial domain. As we are interested in the effects of the deviations of the near-horizon geometry from the Schwarzschild metric on the QNM, we
 employ a rational function approximation given in terms of a two-point Padé approximation \cite{MM:76,S:80,BGM:96,CPVWJ:08} that interpolates between the near-horizon metric \eqref{met} and the asymptotically flat Schwarzschild metric in the far region \eqref{far}. The two-point Padé approximation has seen application a number of scenarios where two complementary asymptotic expansions are available, as in certain classes of supersymmetric Yang-Mills theory \cite{BT:13}, the study of critical phenomena \cite{G:20}, stationary state problems in quantum mechanics \cite{LW:02}, and many other examples where a strong/weak coupling duality is present. Its use in study of the QNM various forms of the Pad\'{e} approximation are discussed in Refs.~\cite{RZ:14,MO:17,H:20,KR:20,K:23}.

We summarize the necessary information about the scheme we use in  Sec.~\ref{p-sec} and describe the resulting radial wave equation in Sec.~\ref{eq-sec}.
Appendix \ref{appC} provides more details and comparisons with other schemes used in the literature. Refs.~\cite{BGM:96,CPVWJ:08}  offer an exhaustive review of all the mathematical aspects of the technique.

\subsection{Pad\'{e} approximants}\label{p-sec}
Two-point Pad\'{e} approximants (finite-order elements of the Pad\'{e} expansion), are designed to approximate a function $f(z)$ that is given by two power series about two different points. If the expansions around $z=0$ and $z=\infty$ of the function are given by
\be\label{l1}
f(z)=\sum_{j=0}^{\infty} c_j z^j, \quad c_j \in \mathbb{C}, \quad c_0 \neq 0\ ,
\ee
\be\label{l2}
f_{\infty}(z)=-\sum_{j=1}^{\infty} c_{-j} z^{-j}, \quad c_{-j} \in \mathbb{C}, \quad c_{-1} \neq 0\ ,
\ee
then a natural continued fraction form known as the M-fraction \cite{MM:76,CPVWJ:08} may be given as
\be\label{mfrac}
M_f(z)= \frac{F_1}{1+G_1 z}+\ContFracOp_{m=2}^{\infty}\left(\frac{F_m z}{1+G_m z}\right)\ ,
\ee
where $F_m \in \mathbb{C} \backslash\{0\}$ and $G_m \in \mathbb{C}$. The operator $\ContFracOp$ represents a continued fraction,
\be
\ContFracOp_{m=1}^{\infty}\left(\frac{a_m}{b_m}\right)=
\cfrac{a_1}{b_1+\cfrac{a_2}{b_2+\cfrac{a_3}{b_3+\cdots}}} \ .
\ee
In practice the fraction is terminated at the $n$-th order, with $a_m=b_m=0$ for $m>n$.
The coefficients in \eqref{mfrac} are given by
\begin{align}
	F_m & =\frac{-H_m^{(-m+1)} H_{m-2}^{(-m+2)}}{H_{m-1}^{(-m+2)} H_{m-1}^{(-m+1)}}\ , \\
	G_m & =\frac{-H_m^{(-m+1)} H_{m-1}^{(-m+1)}}{H_{m-1}^{(-m+2)} H_m^{(-m)}}\ ,
\end{align}
 defined through the Hankel determinants of the coefficients in the original asymptotic expansions \eqref{l1}-\eqref{l2},
\be
\quad H_j^{(m)}:=\begin{array}{|cccc|}
	c_m & c_{m+1} & \cdots & c_{m+k-1} \\
	c_{m+1} & c_{m+2} & \cdots & c_{m+j} \\
	\vdots & \vdots & & \vdots \\
	c_{m+j-1} & c_{m+j} & \cdots & c_{m+2 j-2}
\end{array}
\ee
with $H_0^{(m)}\equiv1$. The truncation of the nesting at some finite $m=n$ results in the coincidence of the expansion of \eqref{mfrac} about $z=0$ and $z=\infty$ with a finite number of terms in Eqs.~\eqref{l1} and \eqref{l2}. A standard procedure allows for the representation of the $n$-th approximant as a rational fraction.

The well-known Rezolla-Zhidenko parameterization \cite{RZ:14} is based on a different family of Pad\'{e} approximants known as C-fractions. Their scheme is both philosophically and technically distinct from our own and we summarize these differences in Appendix~\ref{appC}, while also presenting comparisons for some simple families of metrics at the end of this Section.

Clearly, any such rational function approximation has the potential to develop poles for certain choices of the parameters where the denominator vanishes, and the precise location of the poles will differ with the order of interpolation and number of terms included in the near and far expansions. These must be avoided if they occur outside of the horizon as it makes such expressions unsuitable for use in Eq.~\eqref{kg1}. We thus look for the minimal approximations which focus on the leading  parameters $\alpha_1$ and $\alpha_2$ of the near-horizon expansion and are free of poles in a sufficiently large parameter space.

 We also require that the approximation reduces to the Schwarzschild metric when the coefficients $\alpha_1$ and $\alpha_2$ take on their Schwarzschild values. This does not occur for certain otherwise pole-free Pade\'{e} approximations. Hence we consider four terms in the near-horizon expansion
\be\label{nearexp}
f(r)=\sum_{m=1}^4 \alpha_m\left(\frac{\tilde x}{r_g}\right)^m+\cO(\tilde x^5)  \ ,
\ee
which includes two subleading terms at orders $\tilde x^3$ and $\tilde x^4$ in the near-horizon expansion. This is the minimal number of additional terms required to construct a two-point Padé approximation which is well-behaved in the limit as $\alpha_1\rightarrow 1 $ and $\alpha_2\rightarrow -1$ and  does not develop poles in the region exterior to the horizon near this limit, at least for a finite domain in the parameter space $(\alpha_1,\alpha_2)$.

As we are primarily interested in the effects of small deviations of the metric from the Schwarzschild geometry, we fix three of the parameters to their Schwarzschild values
\be
\alpha_3=1 \ , \quad \alpha_4=-1 \ , \quad M=r_g/2 \quad (\epsilon=0) \label{sch-par} \ .
\ee
 There is no a priori reason to expect that if the parameters $(\alpha_1,\alpha_2)$ in Eq.~\eqref{nearexp} are different from their Schwarzschild counterparts by some $\delta_a$ (i.e. $|\alpha_1 -1|\sim|\alpha_2 +1|\sim\delta_a$) that the rest of the parameters $(\alpha_3,\alpha_4,\epsilon)$ can be taken equal to their Schwarzschild values. In fact, this is not true in many models. On the other hand, the results of Sec.~\ref{results} show that $\delta_a\lesssim 10^{-2}$ already leads to  differences between the QNM frequencies relative to those of the Schwarzschild metric of order $\Delta\omega\sim\mathcal{O}(1)$. However, taking the remaining parameters $(\alpha_3,\alpha_4,\epsilon)$ to differ from their Schwarzschild values also on the order of $\delta_a$, we find that the change in frequency is at most on the order of $\Delta\omega\times\delta_a$, justifying our approximation for the level of precision presented here.

\begin{figure}[ht]
	\begin{center}
		\includegraphics[scale=0.7]{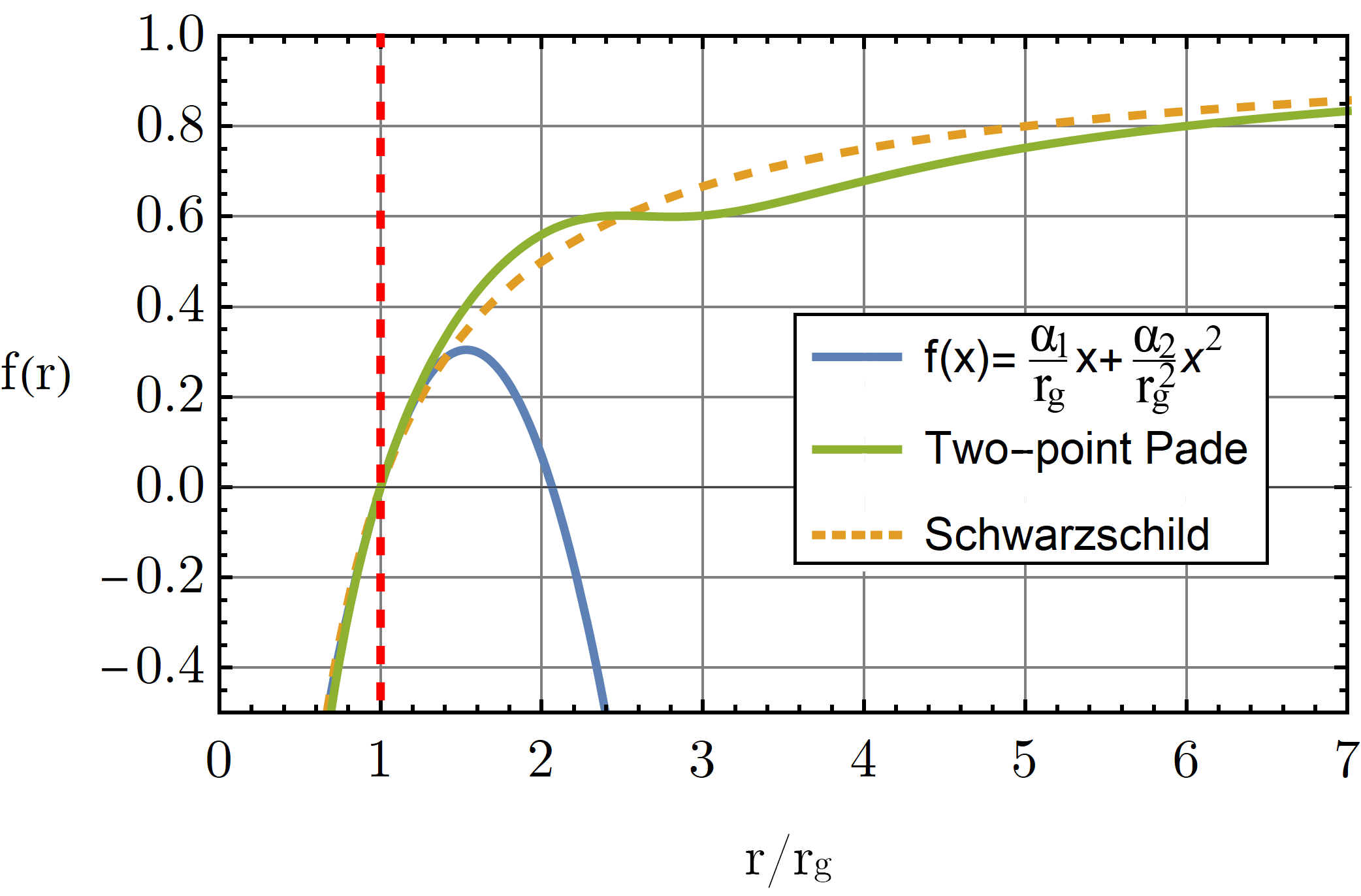}\hspace{20pt}
		\caption{Two-point Padé interpolation \eqref{metPadé} of the near-horizon metric \eqref{nearexp} and the asymptotic Schwarzschild geometry \eqref{far}  with $\alpha_1=1.145$ and $\alpha_2=-1.076$. The horizon is located at the red dashed line.}\label{Padé1}
	\end{center}
\end{figure}

\begin{figure}[ht]
	\begin{center}
		\!\!\!\!\!\includegraphics[scale=0.73]{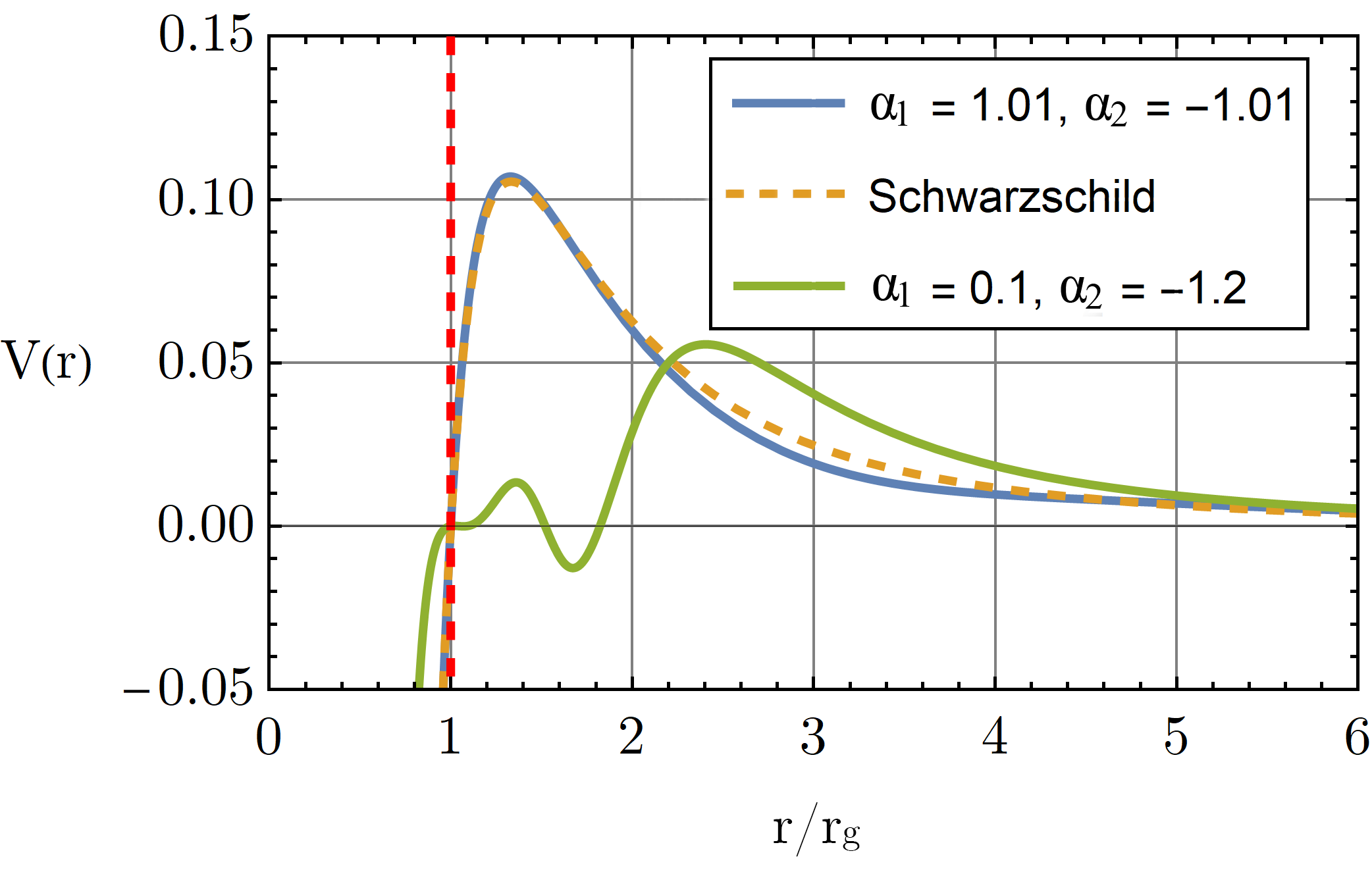}\hspace{20pt}
		\caption{Behaviour of the effective potential as the coefficients $\alpha_1$ and $\alpha_2$ in \eqref{met} deviate from their Schwarzschild values.}\label{pot1}
	\end{center}
\end{figure}

\begin{figure}[ht]
	\begin{center}
		\includegraphics[scale=0.7]{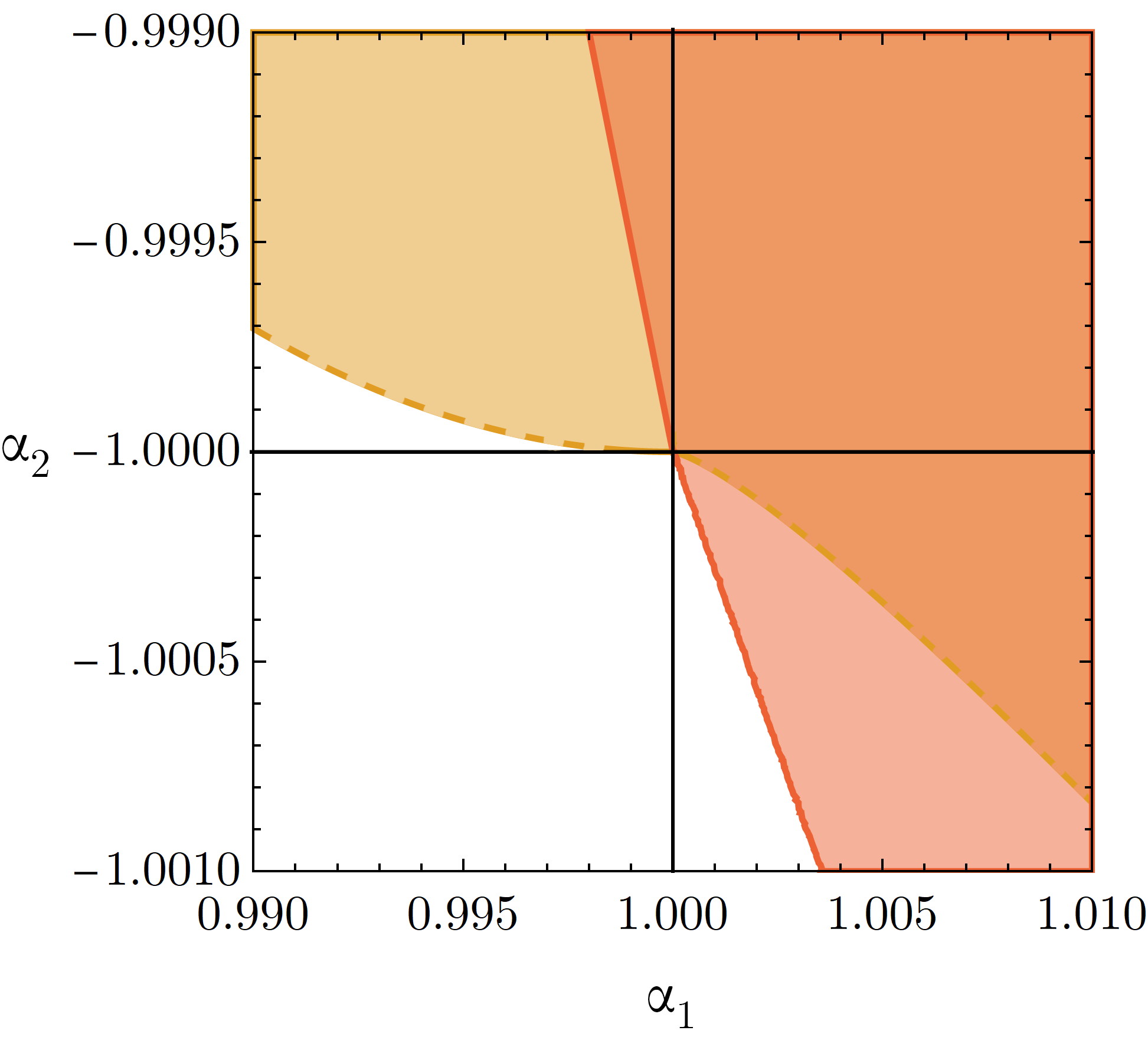}\hspace{16pt}
		\caption{The parameter space $(\alpha_1,\alpha_2)$ covered by the two-point Padé approximation $f_5(r)$ (yellow) and $f_4(r)$ (orange). It is assumed that $\alpha_3=1,\alpha_4=-1$ and $\epsilon=0$. Poles on the interval $r\in(r_g,\infty)$ develop for coefficients lying in the shaded region. The Bardeen metric has $\alpha_2\approx \tfrac{3}{2}-\tfrac{5}{2}\alpha_1$ with $\alpha_1\leqslant 1$ while the static Hayward metric has $\alpha_2\approx \tfrac{5}{2}-\tfrac{7}{2}\alpha_1$ with $\alpha_1\leqslant 1$. The weak energy condition may be satisfied  for $\alpha_1\leqslant1$. The null energy condition is violated for $\alpha_2<-1$.}\label{Padé2}
	\end{center}
\end{figure}

Taking $n=5$ and rationalizing the M-fraction, we obtain the rational function approximation
\be\label{metPadé}
f_5(r)=\frac{Ar^5 +B r^4 +Cr^3 +D r^2 +Er +F}{A r^5 +\tilde{B}r^4 +\tilde{C}r^3 +\tilde{D}r^2 +\tilde{E}r +\tilde{F}}\ ,
\ee
where the coefficients of the numerator are 
\begin{align*}
	A&=\left(3 \;\!\alpha_1^2  +2 \;\!\alpha_1 \:\!\alpha_2  -4 \;\!\alpha_1-3 \;\!\alpha_2 \right)\\
	B&=\left(21 \;\!\alpha_1 r_g-17 \;\!\alpha_1^2 r_g-12 \:\!\alpha_2 \;\!\alpha_1 r_g+16 \:\!\alpha_2 r_g\right)\\
	C&=\left(37 \;\!\alpha_1^2 r_g^2-43 \;\!\alpha_1 r_g^2+28 \;\!\alpha_1 \:\!\alpha_2 r_g^2-33 \;\!\alpha_2 r_g^2+r_g^2\right) \\
	D&= \left(44 \;\!\alpha_1 r_g^3-39 \;\!\alpha_1^2 r_g^3-32 \;\!\alpha_1 \:\!\alpha_2 r_g^3+34 \;\!\alpha_2 r_g^3-3 r_g^3\right)\\
	E&=\left(20 \;\!\alpha_1^2 r_g^4-22 \;\!\alpha_1 r_g^4+18 \;\!\alpha_1 \:\!\alpha_2 r_g^4-18 \:\!\alpha_2 r_g^4+3 r_g^4\right)\\
	F&=(4 \;\!\alpha_1 r_g^5-4 \;\!\alpha_1^2 r_g^5-4 \;\!\alpha_1 \:\!\alpha_2 r_g^5+4 \:\!\alpha_2 r_g^5-r_g^5)
\end{align*}
and of the denominator are
\begin{align*}
	\tilde{B}&=\left(17 \;\!\alpha_1 r_g-14 \;\!\alpha_1^2 r_g-10 \;\!\alpha_2 \;\!\alpha_1 r_g+13 \:\!\alpha_2 r_g\right)\\
	\tilde{C}&=\left(26 \;\!\alpha_1^2 r_g^2-30 \;\!\alpha_1 r_g^2+20 \;\!\alpha_1 \:\!\alpha_2 r_g^2-23 \;\!\alpha_2 r_g^2+r_g^2\right) \\
	\tilde{D}&=\left(27 \;\!\alpha_1 r_g^3-24 \;\!\alpha_1^2 r_g^3-20 \;\!\alpha_1 \:\!\alpha_2 r_g^3+21 \;\!\alpha_2 r_g^3-2 r_g^3\right) \\
	\tilde{E}&=\left(11 \;\!\alpha_1^2 r_g^4-12 \;\!\alpha_1 r_g^4+10 \;\!\alpha_1 \:\!\alpha_2 r_g^4-10 \;\!\alpha_2 r_g^4+2 r_g^4\right)\\
	\tilde{F}&=(2 \;\!\alpha_1 r_g^5-2 \;\!\alpha_1^2 r_g^5-2 \;\!\alpha_1 \:\!\alpha_2 r_g^5+2 \:\!\alpha_2 r_g^5)
\end{align*}
The algorithms for obtaining this expansion are described in \cite{CPVWJ:08}. Appendix \ref{appC} illustrates the relationship between coefficients in the continued fraction representation and the rationalized function for $n=3$. The expansion of $f(r)$ as given by \eqref{metPadé} about $r=r_g$ and $r=\infty$ can easily be verified to coincide with \eqref{nearexp} and \eqref{far} where the parameters are given by Eq.~\eqref{sch-par}. Lower-order ($n<4$) approximations either do not agree in their expansions up to the given order or develop poles on the interval $r\in(r_g,\infty)$ making them ill-suited for numerical integration. The $n=5$ approximation is chosen for its convenient domain of validity in the parameter space of $(\alpha_1,\alpha_2)$. In \figurename{\ref{Padé1}} and \figurename{\ref{pot1}} we show, respectively, the interpolating function along with the effective potential as they deviate from the Schwarzschild limit for some values of $\alpha_1$ and $\alpha_2$. ~\figurename{\ref{Padé2}} shows the domain in the $(\alpha_1,\alpha_2)$ plane of parameters for which both the $n=4$ and $n=5$  Pad\'{e} approximations are pole-free.

\begin{figure}[ht]
	\begin{center}
		\caption{Parameter space  $(\alpha_1,\alpha_2)$ as covered by  the two-point Padé approximation using the Rezzolla--Zhidenko scheme (Appendix~\ref{appC}  with $n=2$  (yellow) and $n=3$ (orange). It is assumed that $\alpha_3=1,\alpha_4=-1$ and $\epsilon=0$. Poles on the interval $r\in(r_g,\infty)$ develop for coefficients lying in the shaded region.  The weak energy condition may be satisfied  for $\alpha_1\leqslant1$. The null energy condition is violated for $\alpha_2<-1$.}\label{RZ}
	\end{center}
\end{figure}

The rational function approximation above is also the highest-order approximation for which the tortoise coordinate is analytically integrable, though we will instead work with an expansion of the tortoise coordinate due to the verbose nature of the exact solution (it contains $>1500$ terms).

It is well-known that certain modifications to the Regge--Wheeler potential impart significant modifications to the QNM spectrum, and one may worry that such an interpolation will therefore lead to erroneous results. In the work of Nollert \cite{N:96}, it was shown that approximating the Regge--Wheeler potential via piece-wise step functions leads to drastic changes in the QNM spectrum. Subsequent studies have shown that piecewise continuous approximations enable one to compute the ringdown signal to arbitrary precision, however the QNM spectrum remains modified relative to the Schwarzschild case \cite{D:20}. It is precisely this high sensitivity to small changes in the potential which make QNMs powerful observational probes of deviations from the classical Kerr and Schwarzschild solutions. In the present case, it is important that only the deviations encoded in the semi-classical expansion coefficients affect the QNM spectrum, and that it is not artificially modified by the interpolation scheme. Thus, we will also consider the $n=4$ interpolation
\be\label{metPadé2}
f_4(r)=\frac{Ar^4 +B r^3 +Cr^2 +D r +E}{A r^4 +\tilde{B}r^3 +\tilde{C}r^2 +\tilde{D}r +\tilde{E} }\ ,
\ee
where the coefficients are given by
\begin{align*}
	A&=(\alpha_1^2-3 \:\!\alpha_1-2\:\! \alpha_2)\\
	B&=(r_g-5\:\! \alpha_1^2 r_g+13\:\! \alpha_1 r_g+9\:\! \alpha_2 r_g)\\
	C&=(9\:\! \alpha_1^2 r_g^2-20\:\! \alpha_1 r_g^2-14 \:\!\alpha_2 r_g^2-3 \:\!r_g^2) \\
	D&= (3\:\! r_g^3-7 \:\!\alpha_1^2 r_g^3+14\:\! \alpha_1 r_g^3+9 \:\!\alpha_2 r_g^3)\\
	E&=(2\:\! \alpha_1^2 r_g^4-4 \:\!\alpha_1 r_g^4-r_g^4-2 \:\!\alpha_2 r_g^4) \\
	\tilde{B}&=(r_g-4 \:\!\alpha_1^2 r_g+10 \:\!\alpha_1 r_g+7\:\! \alpha_2 r_g)\\
	\tilde{C}&=(6\:\! \alpha_1^2 r_g^2-13\:\! \alpha_1 r_g^2-9 \:\!\alpha_2 r_g^2-2\:\! r_g^2)\\
	\tilde{D}&= (2\:\! r_g^3-4\:\! \alpha_1^2 r_g^3+8\:\! \alpha_1 r_g^3+5\:\! \alpha_2 r_g^3)\\
	\tilde{E}&= (\alpha_1^2 r_g^4-2\:\! \alpha_1 r_g^4+\alpha_2 r_g^4)\ ,
\end{align*}
and compare the relative size of the difference between the two functions for a given choice of $\alpha_1$ and $\alpha_2$ with the deviation in frequency induced by those values. For the interpolating function $f_4(r)$, poles appear for different values of the coefficients compared to $f_5(r)$. Computation using both interpolations for the parameters $(\alpha_1,\alpha_2)$ belonging to the overlapping pole-free domains indicate that the results differ only by order $\delta_a^2$, and thus the $\mathcal{O}(1)$ deviations observed when $(\alpha_1,\alpha_2)$ vary are not an artifact of the choice of interpolation order.

\subsection{The wave equation}\label{eq-sec}

Equation \eqref{radial1} along with the boundary conditions of Eqs. \eqref{bcin}-\eqref{bcout} furnish a standard perturbation problem in spherical symmetry. To solve it, we construct an ansatz which has the correct behaviour at the boundaries. Near the horizon the tortoise coordinate takes the form
\begin{align}
	r_*&=\int\!\dfrac{dr}{f(r)}{\approx}\frac{r_g}{\alpha_1}\log (r-r_g)-\dfrac{\alpha_2 (r-r_g)}{\alpha_1^2}\ ,
\end{align}
where only terms up to order $\mathcal{O}(\tilde x)$ are kept. Taking the limit \eqref{schwlim}, the tortoise coordinate reduces to the known form
\be
r_*=r+r_g\log\left(\frac{\tilde x}{r_g}\right)\ .
\ee
Near the horizon we therefore seek purely ingoing solutions of the form
\begin{align}
	\phi\sim e^{-i\omega r_*}&=\left(r-r_g\right)^{-\frac{i w}{a_1}}e^{-i w \chi}R(r)\ ,
\end{align}
where we have defined
\be
\chi\defeq\-\frac{\alpha_2 (r-r_g)}{\alpha_1^2}\ .
\ee
At infinity, we assume the metric approaches the asymptotically flat Schwarzschild form. The boundary condition at infinity is
\be
\phi\sim e^{i\omega r_*}=e^{i\omega r} \left(\dfrac{r}{r_g}\right)^{\!i\omega r_g}\!\!R(r)\ .
\ee
The following ansatz achieves these limiting forms while possessing the correct limit when the coefficients $a_1$ and $a_2$ take on their Schwarzschild values:
\be
\phi(r)= e^{-i\omega \chi}(r-r_g)^{-\tfrac{i\omega}{a_1}}e^{2i\omega r}\left(\frac{r}{r_g}\right)r^{\tfrac{i\omega}{a_1}}R(r)
\ee
Inserting the ansatz into \eqref{radial1} and using \eqref{met} and \eqref{pot} reduces the radial equation to the form
\be\label{radial2}
R''(r)+\tau(r)R'(r)+\sigma(r)R(r)=0\ ,
\ee
Before determining the coefficients  we transform the radial domain to a compact interval by introducing \cite{RZ:14}
\be
x\defeq 1-\frac{r_g}{r}\ ,
\ee
 so that  $r\in[r_g,\infty)$ is mapped to $x\in[0,1]$. We further let $\tilde{R}(x)=x(1-x)R(x)$ so that the boundary conditions become $\tilde{R}(0)=\tilde{R}(1)=0$. The radial equation \eqref{radial2} becomes
\begin{align}
	\tilde{R}''(x)+\tau(x)\tilde{R}'(x)+\sigma(x)\tilde{R}(x)=0\ ,
\end{align}
where the functions $\{\tau(x),\sigma(x)\}$ are given in Appendix \ref{appB}. When $a_1=1 $ and $a_2=-1$ the radial equation reduces to
\begin{align}
	&\left[\frac{ L^2+L-8 r_g^2 \omega ^2+8 i r_g \omega +3-2 i r_g \omega -1}{(x-1)^3 x}\right.\nonumber\\
	&\left.-\frac{ \left(L^2+L-3 (i-2 r_g \omega)^2\right)+x (i-2 r_g \omega)^2}{(x-1)^3 }\right]\tilde{R}(x)\nonumber\\
	&+\frac{ \left(x^2 (-1-4 i r_g \omega )+x (2+8 i r_g \omega )-2 i r_g \omega -1\right)}{(x-1)^2 x}\tilde{R}'(x)\nonumber\\
	&+\tilde{R}''(x)=0\ ,
\end{align}
which is the known form of the radial equation for scalar quasinormal perturbations on the Schwarzschild background.

\section{Matrix method}\label{matrix}

A variety of numerical and semi-analytic methods exist for computing QNMs, including the Eikonal approximation for $l\gg1$ \cite{P:71,G:72}, the WKB approximation for low harmonic number \cite{SW:85,IW:87}, pseudo-spectral methods \cite{J:17}, and others, with analytical results being available for certain effective potentials (see \cite{BV:11} for a survey). In this work, we implement a modified version of the matrix method described in \cite{LQ:16,LQ:17} to compute the quasinormal frequencies associated to \eqref{radial2}, which offers improved efficiency and flexibility over other finite difference methods. We briefly summarize the approach here. First note that a general master equation for the radial part of the wavefunction can be written
\be
G(x,\omega) \phi(x)=0\ ,
\ee
where $G(x,\omega)$ is a differential operator and $x$ is the compact radial coordinate defined previously. Discretization of the differential equation on a  grid of $N$ points $x\in\{x_i,...,x_N\}$ allows one to cast the problem as a homogeneous matrix equation
\be\label{mateq}
\bar{M}(\omega)\phi=0\ ,
\ee
where $\phi$ is a column vector consisting of the function value $\phi(x_i)$ at each grid point $x_i$. The matrix $M$ is determined using Cramer's rule to compute, e.g.
\be
\phi''(x_1)=\dfrac{\text{det}(M^{(x_1)}_2)}{\text{det}{(M)}}
\ee
from the matrix $M$ of coefficients of the Taylor series of $\phi(x)$ at a grid point $x=x_0$. For example, for the point $x=x_0$ one has $\phi=M D$ such that
\be
\left(\!\!\begin{array}{c}
	\phi\left(\delta x_1\right) \\
	\phi\left(\delta x_2\right) \\
	\phi\left(\delta x_3\right) \\
	\cdots
\end{array}\!\!\right)=\left(\!\begin{array}{ccc}
	x_1-x_0 & \frac{\left(x_1-x_0\right)^2}{2} & \frac{\left(x_1-x_0\right)^3}{3 !}   \\
	x_2-x_0 & \frac{\left(x_2-x_0\right)^2}{2} & \frac{\left(x_2-x_0\right)^3}{3 !}  \\
	x_3-x_0 & \frac{\left(x_3-x_0\right)^2}{2} & \frac{\left(x_3-x_0\right)^3}{3 !}   \\
	\ldots & \ldots & \ldots
\end{array}\!\right)\!\!\left(\!\!\begin{array}{c}
	\phi^{\prime}\left(x_0\right) \\
	\phi^{\prime \prime}\left(x_0\right) \\
	\phi^{\prime \prime \prime}\left(x_0\right) \\
	\cdots
\end{array}\!\!\right)\nonumber
\ee
The determinants are computed by first computing the $ij$-minors of each $M$ and then using a using a cofactor expansion. The condition that \eqref{mateq} has non-trivial solutions is then that
\be\label{det}
\text{det}(\bar{M}(\omega))=0
\ee
which gives a polynomial for $\omega$ which can be solved numerically. This method has previously been employed to study scalar QNMs for asymptotically flat and (anti)-de Sitter Schwarzschild black holes \cite{LQ:17}, Kerr and Kerr-Sen black holes \cite{LQ:17b}, and accreting Vaidya black holes with a specific choice of mass function \cite{LS:21}, among others. Compared to other numerical and semi-analytic methods, the matrix method shows improved utility in a number of ways. By allowing for discretization on non-uniform grids, one is able to preferentially increase resolution in the most sensitive parts of the integration domain. Boundary conditions are also straightforwardly implemented by replacing $\bar{M}_{ij}\rightarrow\delta_{ij} \text{ for }i=\{1,N\}$. Note that we refrain from using methods based on double-null constructions since generally dynamical backgrounds do not admit analytic double-null foliations (or the explicit coordinate transformations are not known explicitly).

The matrix method is able to achieve good numerical accuracy with flexible efficiency/precision trade-offs determined primarily by the grid size (and step size, if evolving in time). For example, with a uniformly spaced grid $x_i\in\{0,\tfrac{1}{4},\tfrac{2}{4},\tfrac{3}{4},1\}$, one obtains for the $l=m=0$ fundamental mode of the Schwarzschild black hole:
\be
\omega_{000}=0.21863 - 0.19826 i
\ee
This can be compared to the known frequency obtained through Leaver's method or pseudo-spectral/AIM methods \cite{M:22}:
\be
\omega_{000}=0.22091-0.20979i
\ee
With $x_i\in\{0,\tfrac{1}{20},...,\tfrac{19}{20},1\}$, we obtain instead:
\be
\omega_{000}=0.22088 - 0.20979 i\ .
\ee
Additional comparisons are given in Appendix \ref{appD}.
\\

Our implementation of this method includes an important modification required for our analysis. The condition \eqref{det} which is solved numerically to determine the quasinormal frequency $\omega$ is in general a polynomial of degree $N$, thus possessing $N$ complex-valued solutions. Only one of these corresponds to the quasinormal frequency while the rest are spurious solutions, whose number increases linearly with the precision of the grid. Thus, as more precision is used (in the form of a larger number of grid points) it becomes increasingly difficult to identify the correct frequency. When the quasinormal frequency is already known from other analytic or numerical methods, one can seed a root finding algorithm for \eqref{det} with the known frequency and look for the solution which converges as $N$ is increased. However, we have no prior knowledge of the correct quasinormal frequency for a generic near-horizon metric of the form \eqref{met}.
\\

Thus we modify the algorithm to proceed in a sequence of steps, as follows. Since our interpolating metric \eqref{metPadé} and radial wavefunction \eqref{radial2} possess smooth limits to the Schwarzschild case when $
\alpha_1\rightarrow 1 $ and $\alpha_2\rightarrow -1 $, we proceed in a sequence of steps beginning with this limit, then increment the coefficients $\alpha_1$ and $\alpha_2$ in finite steps $\Delta \alpha_1$ and $\Delta \alpha_2$ until the target values are reached. At each step, the root finder is seeded with the result of the previous step and picks out the solution which minimizes the $L2$ distance from the quasinormal frequency computed in the previous step. In this way the real quasinormal frequency is identified and tracked, which should vary smoothly away from its Schwarzschild value as the interpolating metric does.

\section{Quasinormal modes}\label{results}

 We now compute the low-lying $l=0,1,2$ quasinormal frequencies for the fundamental modes. Note that in spherical symmetry modes of the same $l$ but different $m$ are degenerate. In the absence of observational constraints on scalar perturbations, we assume a hypothetical 10\% bound on the deviations of these frequencies from their Schwarzschild values, and use these to infer bounds and sensitivity on the leading coefficients $(\alpha_1,\alpha_2)$ of \eqref{met} and thus, the leading coefficients $(E,e_2)$ of the EMT expansion near the horizon. For all computations, a sufficiently high number of grid points is chosen to achieve a numerical error of at most $\mathcal{O}(10^{-4})$, below which we expect the choice of interpolation procedure and truncation of the series \eqref{nearexp} to dominate the error.

 In all calculations below we set $r_g=1$.

\subsection{$l=0$}

\begin{figure}[ht]
	\begin{center}
		\includegraphics[scale=0.55]{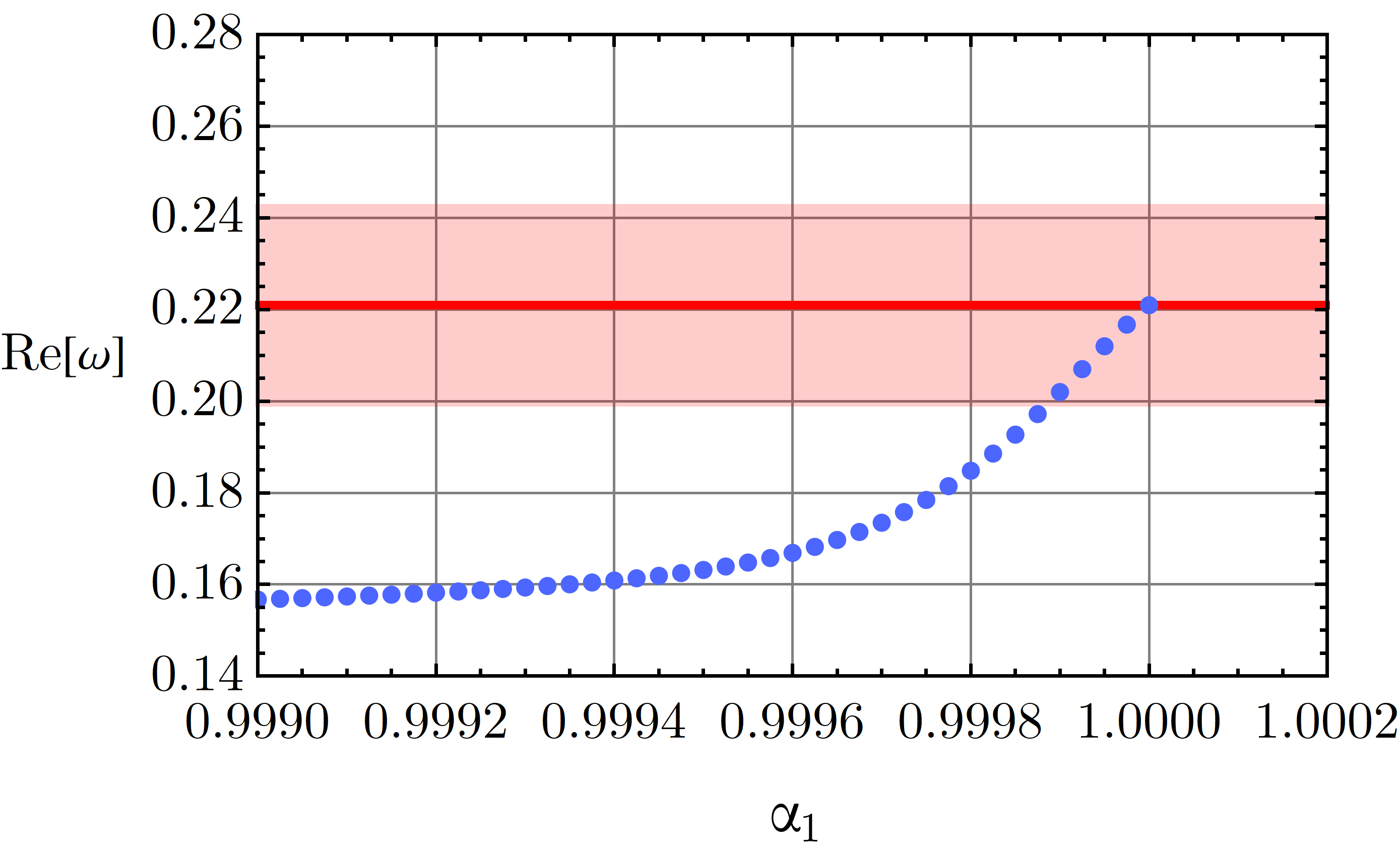}\\	\!\!\!\includegraphics[scale=0.57]{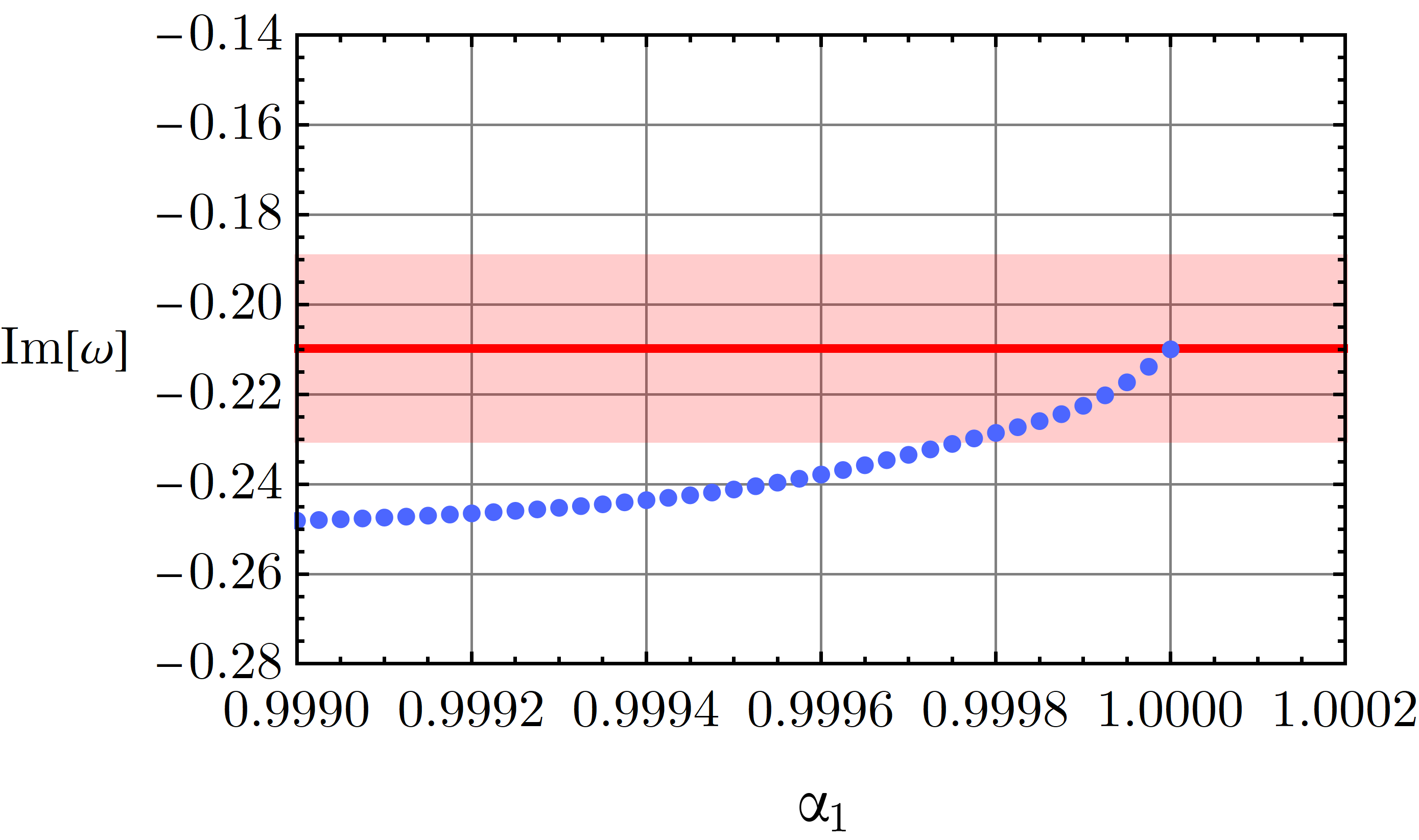}\\
		\!\!\!\includegraphics[scale=0.56]{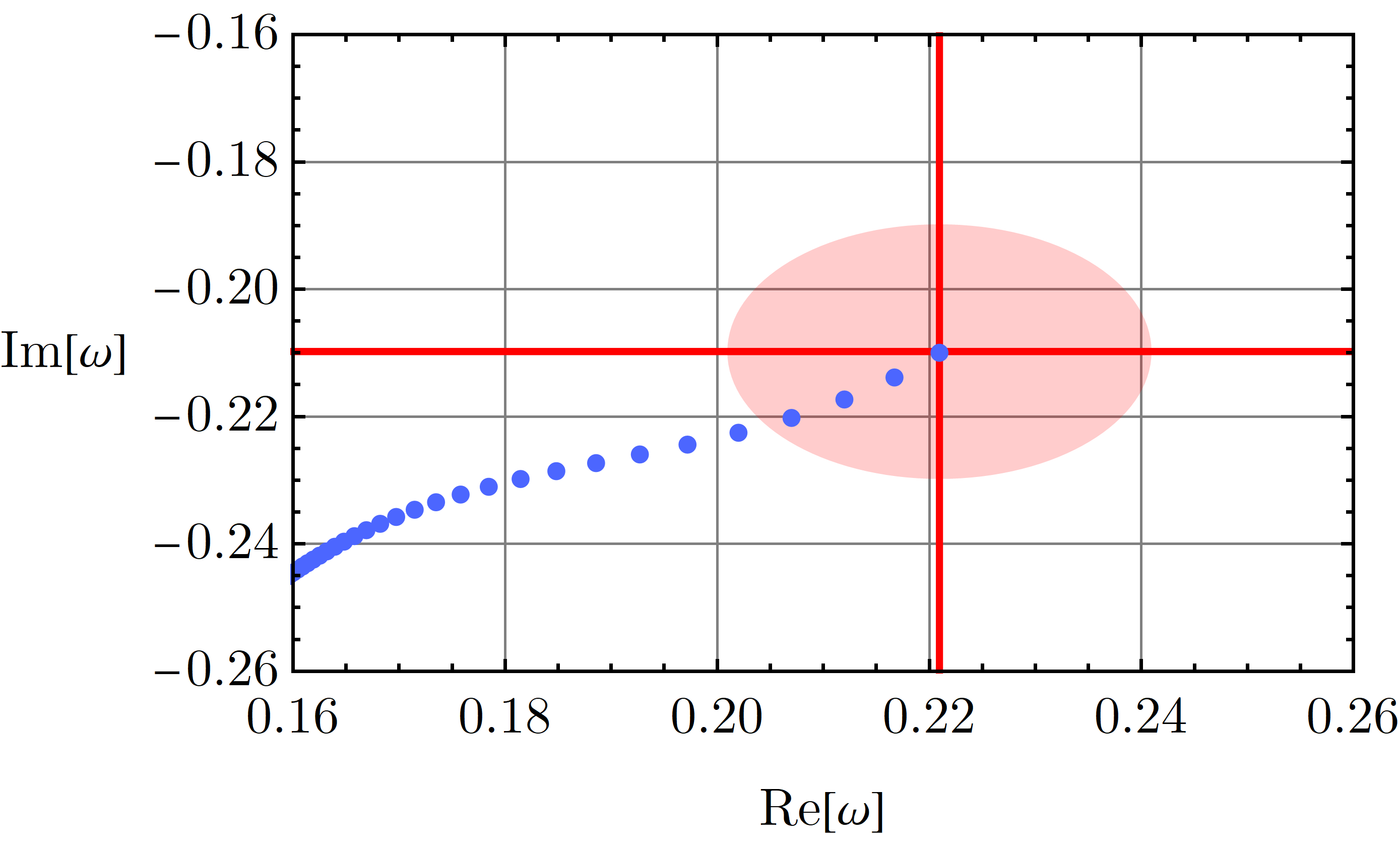}
		\caption{Deviation from GR prediction for the frequency and damping time of (000) as a function of $\alpha_1$ with $\alpha_2=0$. The red shaded region corresponds to a 10\% deviation from the  Schwarzschild value as given in \eqref{con1}. The step size is $\Delta \alpha_1=0.000025$ in all  panels.}\label{l0qnms}
	\end{center}
\end{figure}

\begin{figure}[ht]
	\begin{center}
		\includegraphics[scale=0.97]{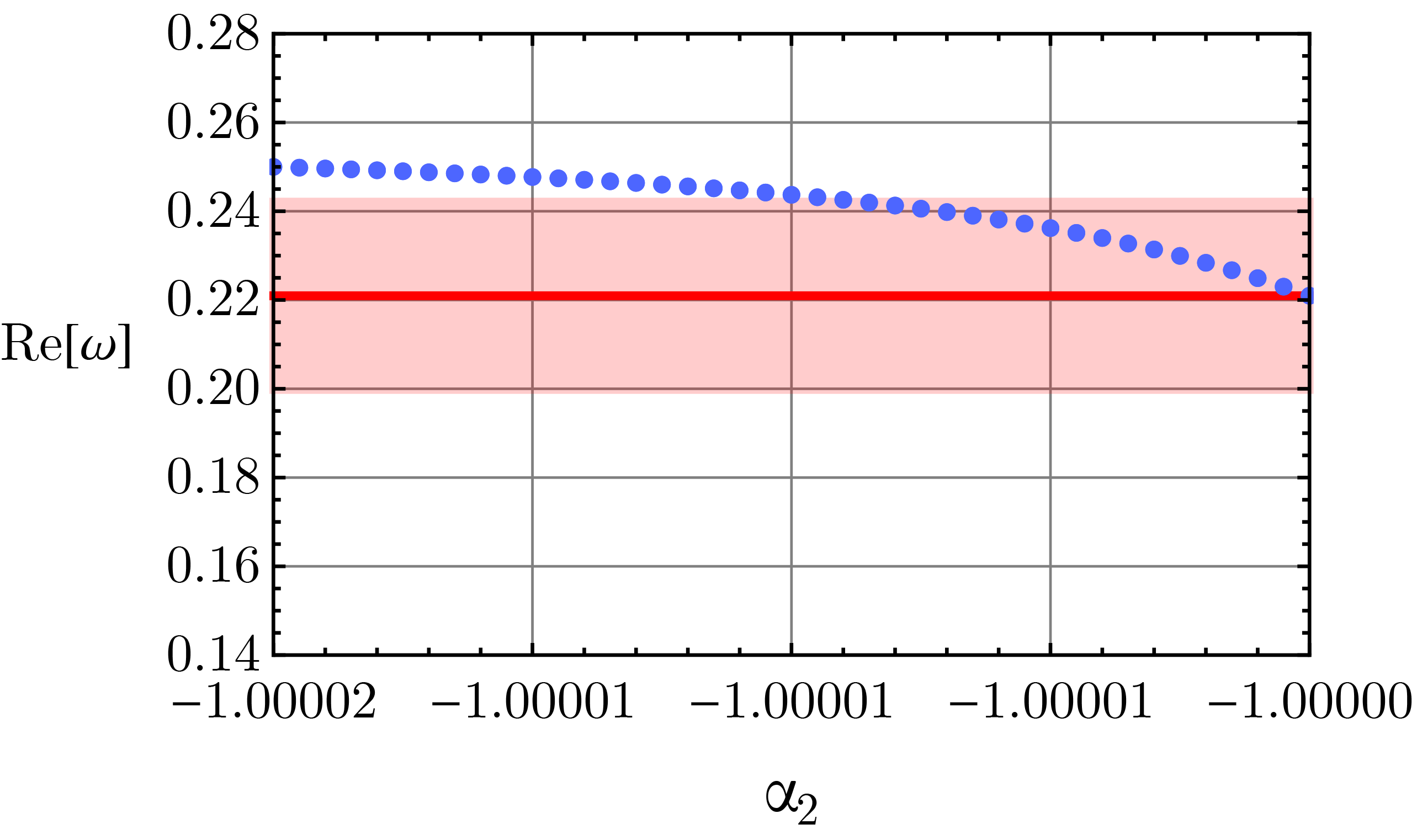}\\	\!\!\!\includegraphics[scale=0.95]{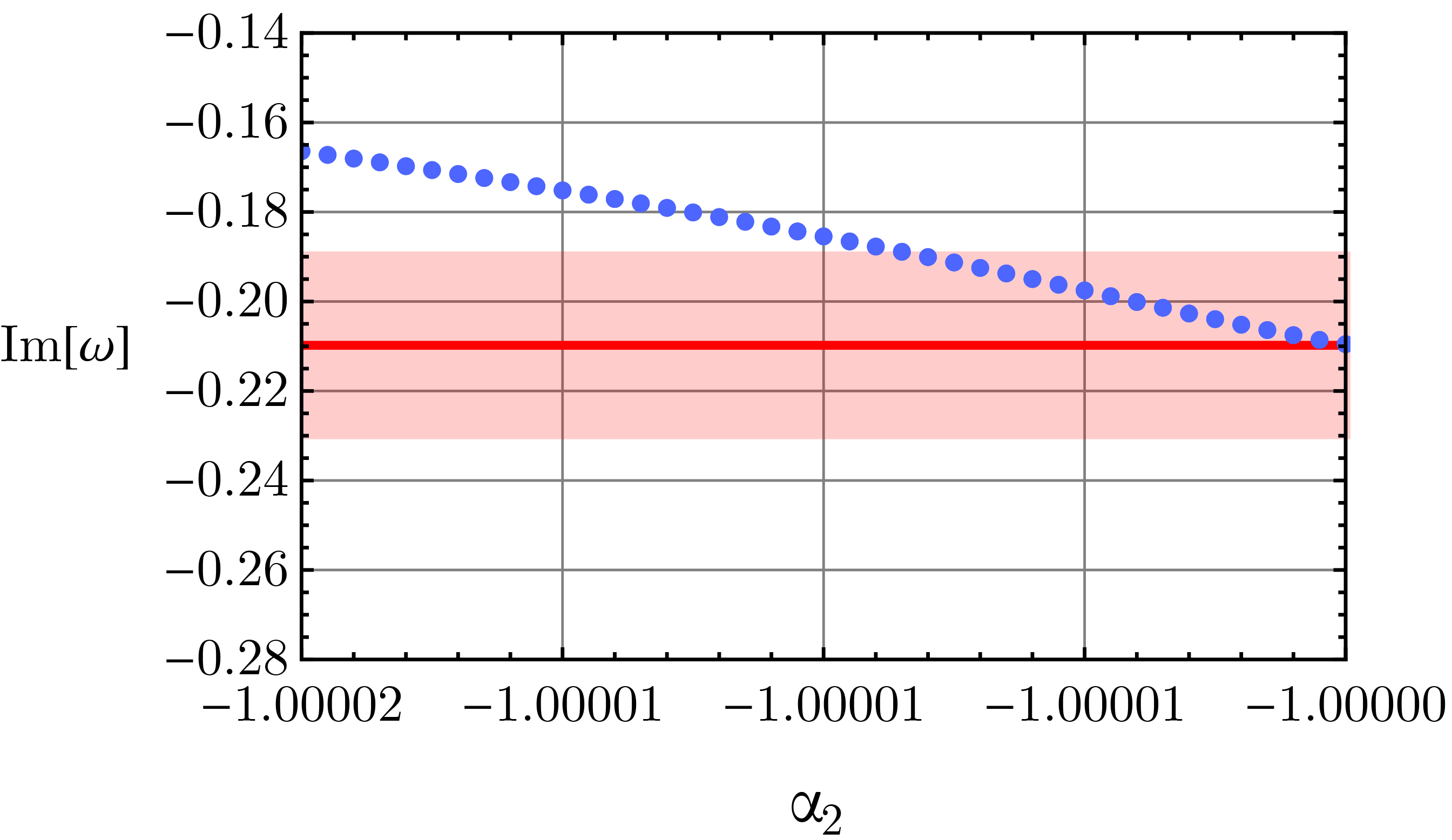}\\
		\!\!\!\!\!\!\!\!\!\!\includegraphics[scale=0.97]{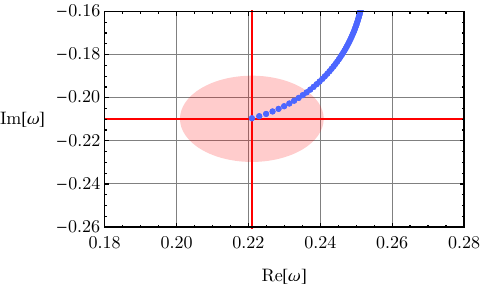}
		\caption{Deviation from GR prediction for the frequency and damping time of (000) as a function of $\alpha_2$ with $\alpha_1=0$. The red shaded region corresponds to a 10\% deviation from the  Schwarzschild value as given in \eqref{con1}. The step size is $\Delta \alpha_2=0.000001$ in all panels.}\label{l0qnmsa2}
	\end{center}
\end{figure}

We compute the frequency and damping time of the (000) mode on a uniformly spaced grid with $N=30$, as a function of $\alpha_1$ with $\alpha_2$ fixed and $\alpha_2$ with $\alpha_1$ fixed. The results are displayed in \figurename{\ref{l0qnms}} and \figurename{\ref{l0qnmsa2}}. For the Schwarzschild metric, the scalar mode has frequency $\omega_{000}=0.220952-0.209795i$, with a 10\% deviation implying
\begin{align}\label{con1}
	&\text{Re}[\omega_{000}]\in(0.198857,0.243047)\ ,\nonumber\\
	&\text{Im}[\omega_{000}]\in(0.188816,0.230775)\ .
\end{align}

In both the frequency and damping time of the $(000)$ mode, fractional deviations from the Schwarzschild frequency on the order $0.1$ occur already with $\Delta \alpha_1=10^{-4}$, showing high sensitivity to the leading coefficient of \eqref{met}. As a check, we also compare the result obtained for $\omega_{000}$ using the interpolating metric \eqref{metPadé} with the value obtained using \eqref{metPadé2} on the same grid. For example, taking $\alpha_1=0.9999$ with $\alpha_2=-1$ we find
\begin{align}
\omega_{000}&= 0.22080-0.21006i  \qquad \text{with}\quad  f_5(r) \\
\omega_{000}&= 0.22167-0.21567i  \qquad \text{with}\quad  f_4(r)\ .
\end{align}
The error introduced by the ambiguity in choice of interpolation order is therefore smaller than the effect induced by deviations from the Schwarzschild metric at leading and subleading order in the near-horizon expansion. We have also verified that a similar agreement occurs for the fundamental $l=1,2$ modes below.

\subsection{$l=1$}

\begin{figure}[ht]
	\begin{center}
		\includegraphics[scale=0.55]{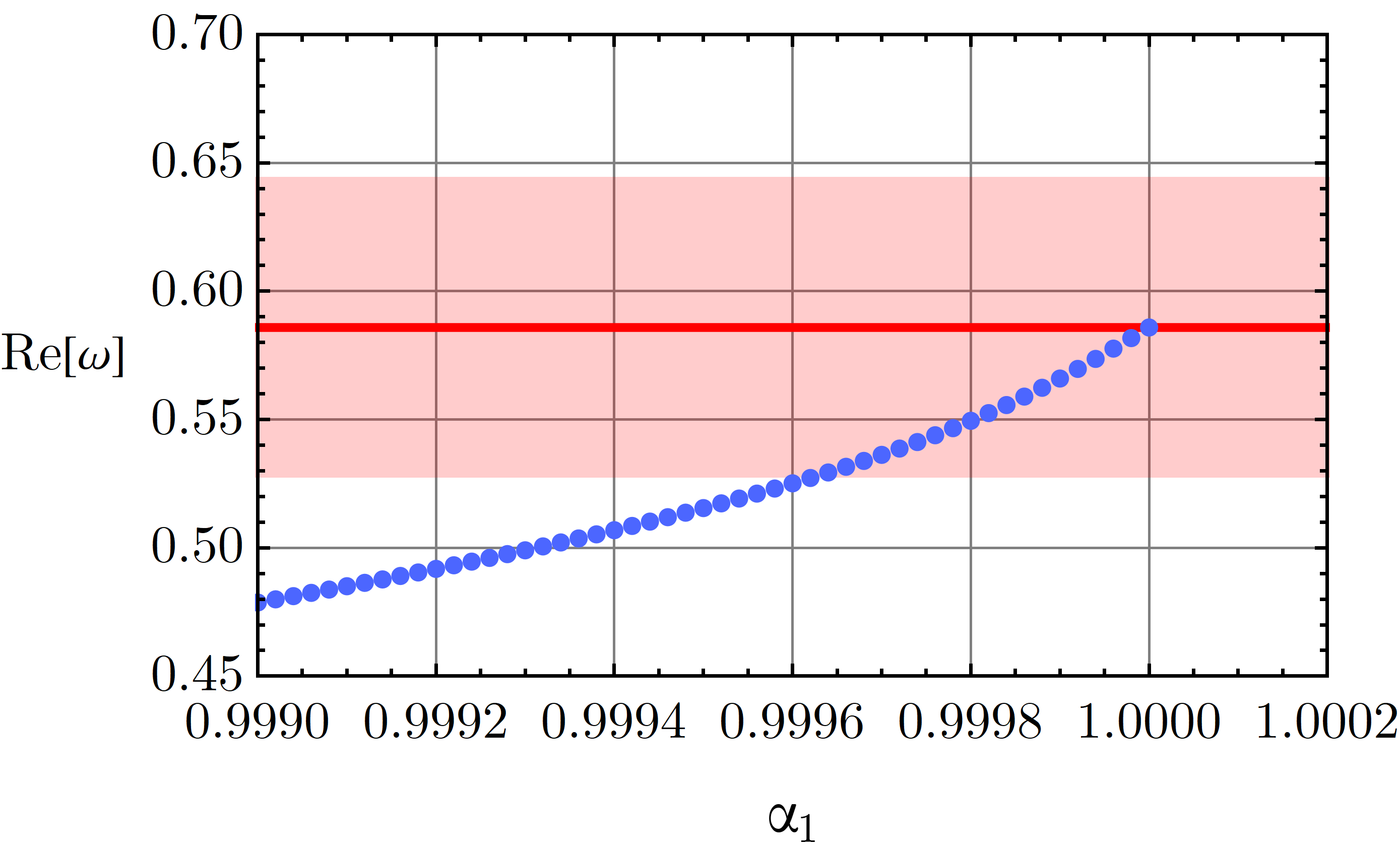}\\	
		\!\!\!\!\includegraphics[scale=0.57]{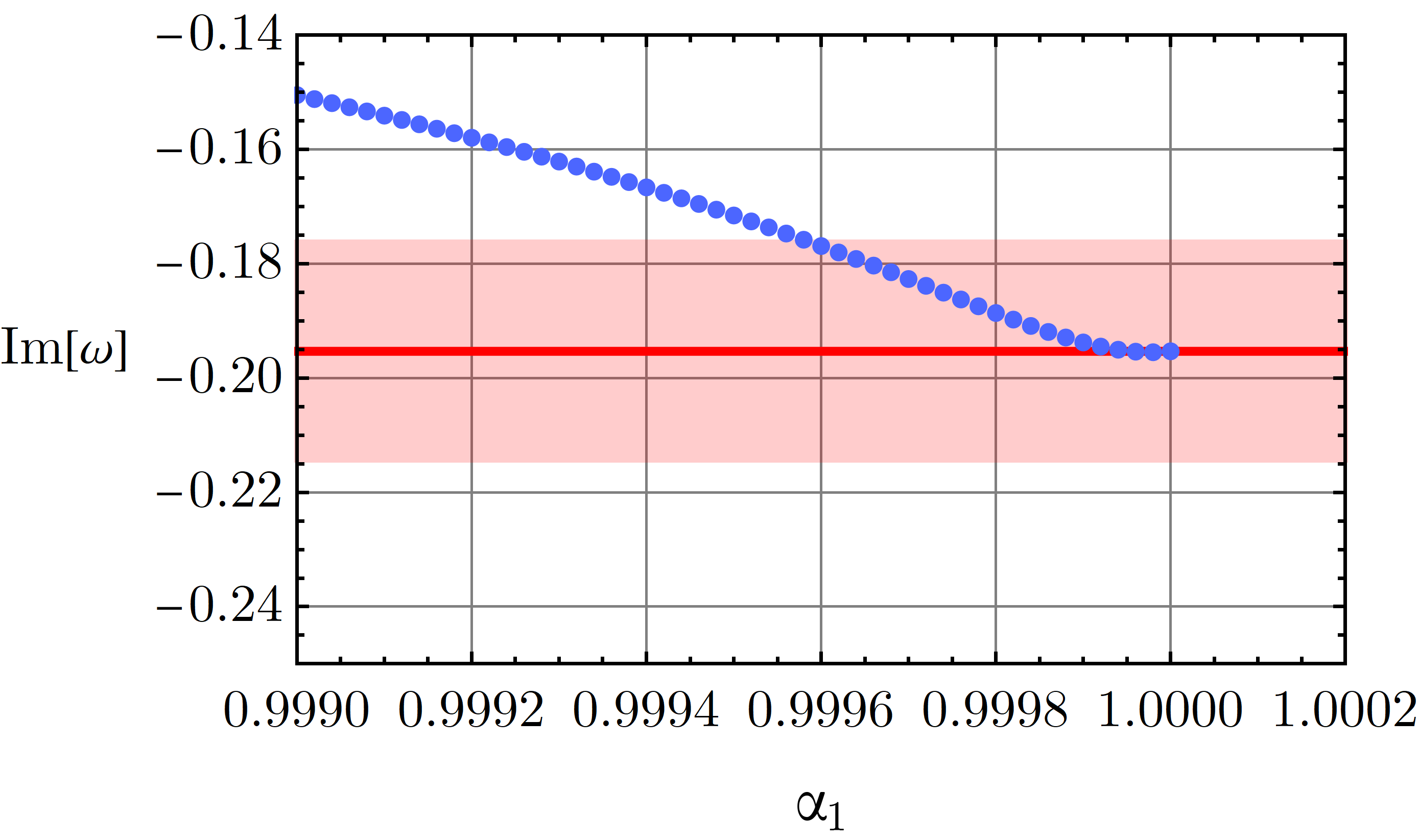}\\
		\!\!\!\!\!\includegraphics[scale=0.55]{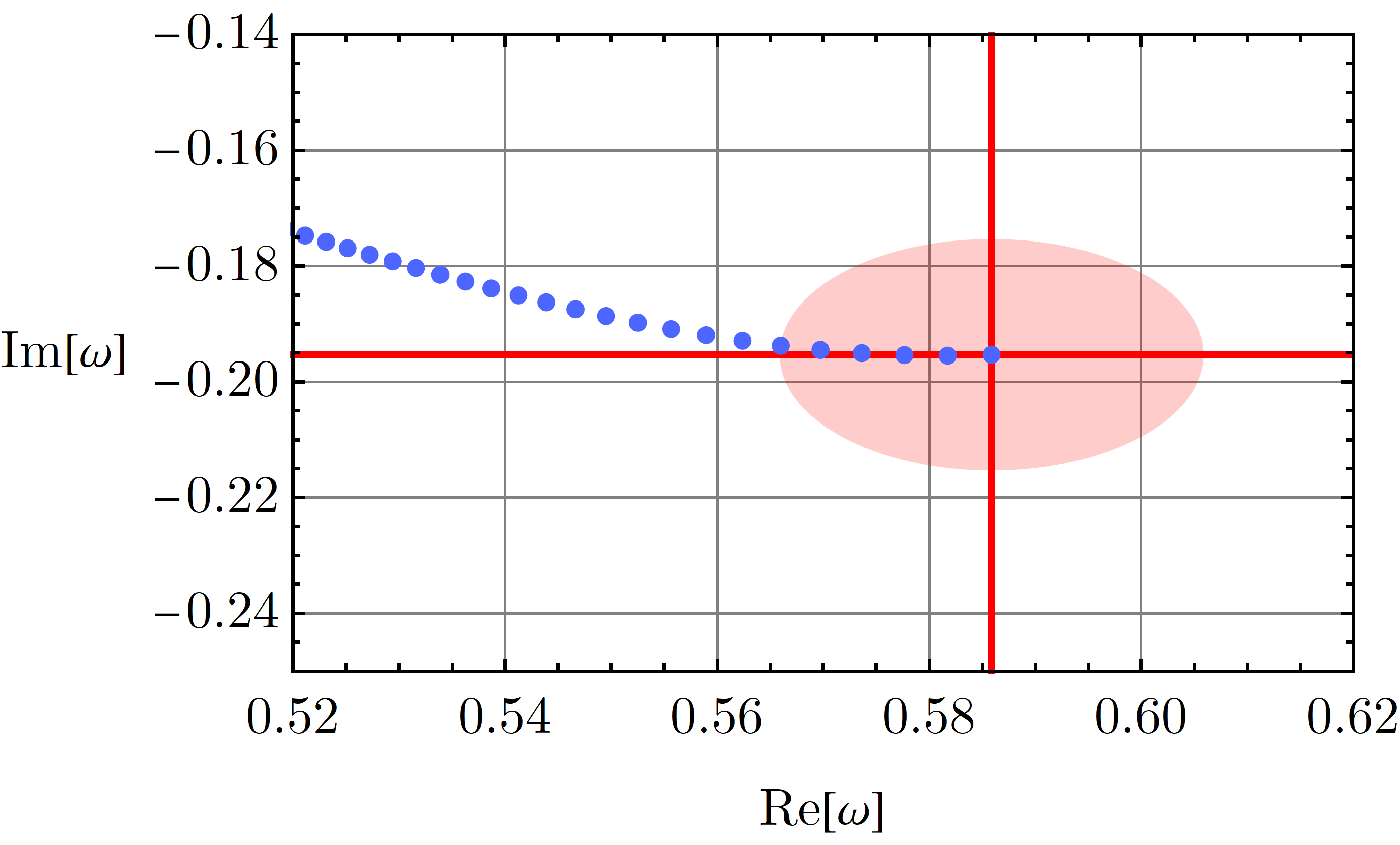}
		\caption{Deviation from GR prediction for the frequency and damping time of (100) as a function of $\alpha_1$ with $\alpha_2=0$. The red shaded region corresponds to a 10\% deviation from the  Schwarzschild value as given in \eqref{con1}. The step size is $\Delta a_1=0.00002$ in all panels.}\label{l1qnms}
	\end{center}
\end{figure}

\begin{figure}[ht]
	\begin{center}
		\includegraphics[scale=0.56]{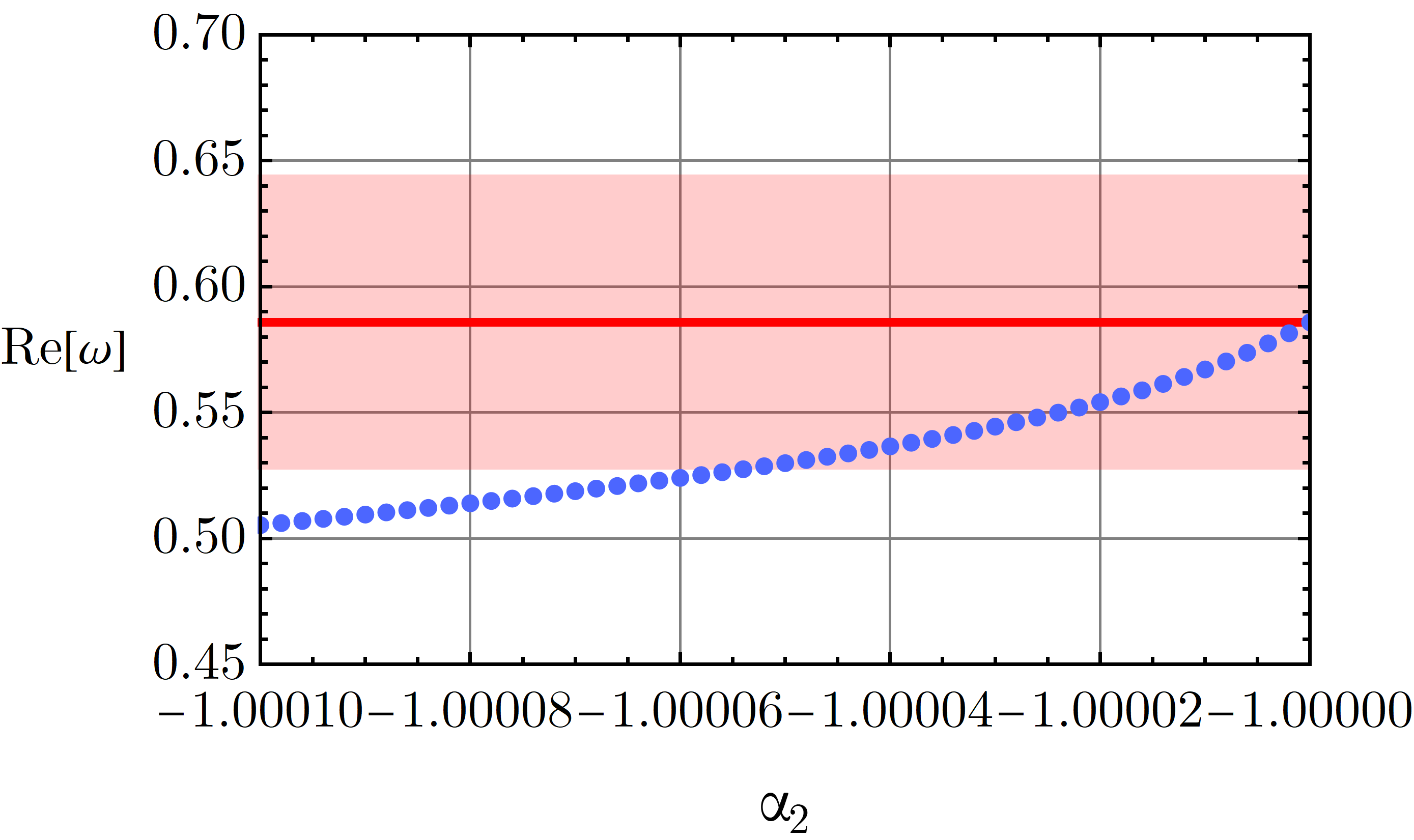}\\	\!\!\!\includegraphics[scale=0.58]{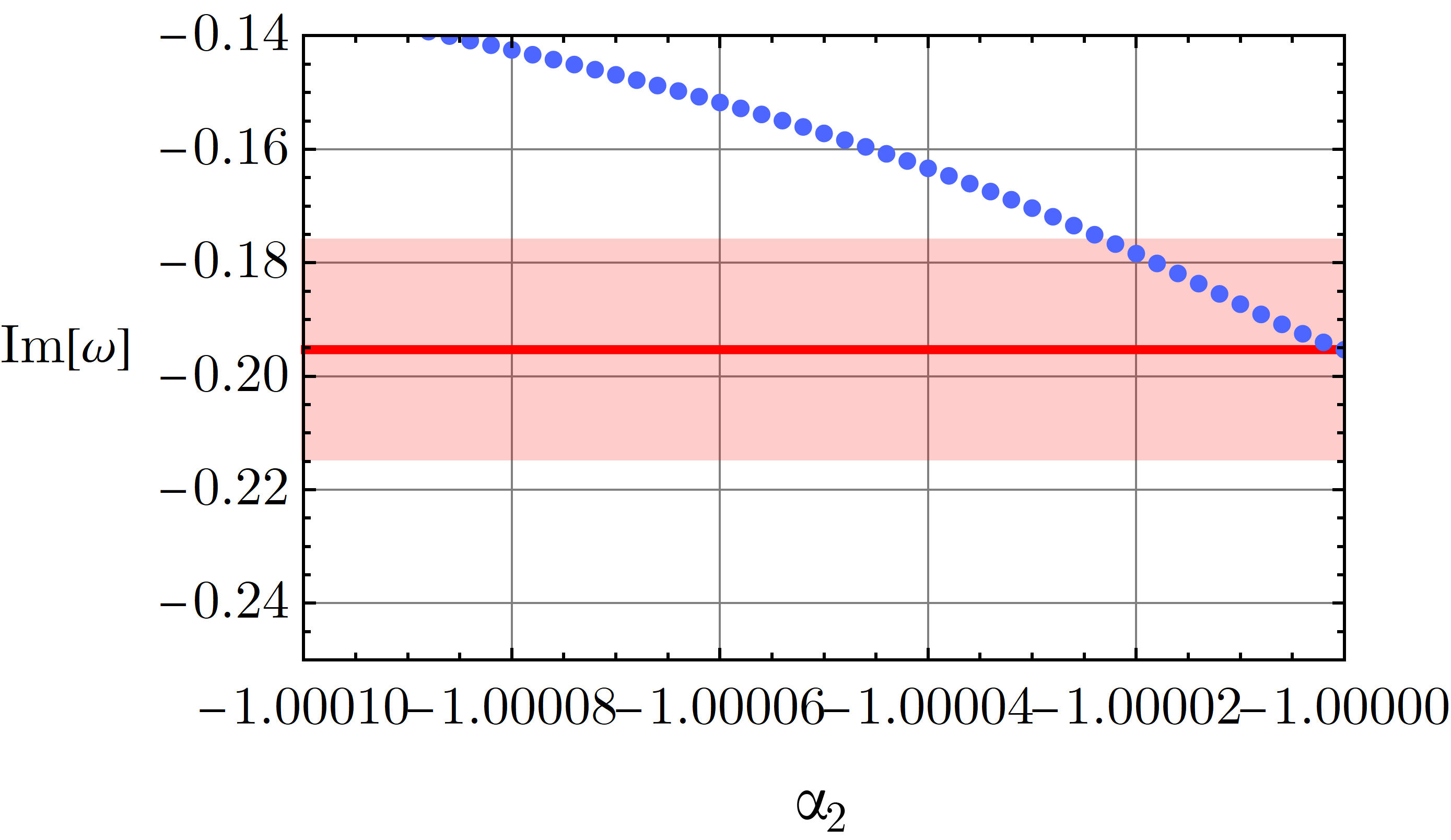}\\
		\!\!\!\!\!\!\!\!\includegraphics[scale=0.55]{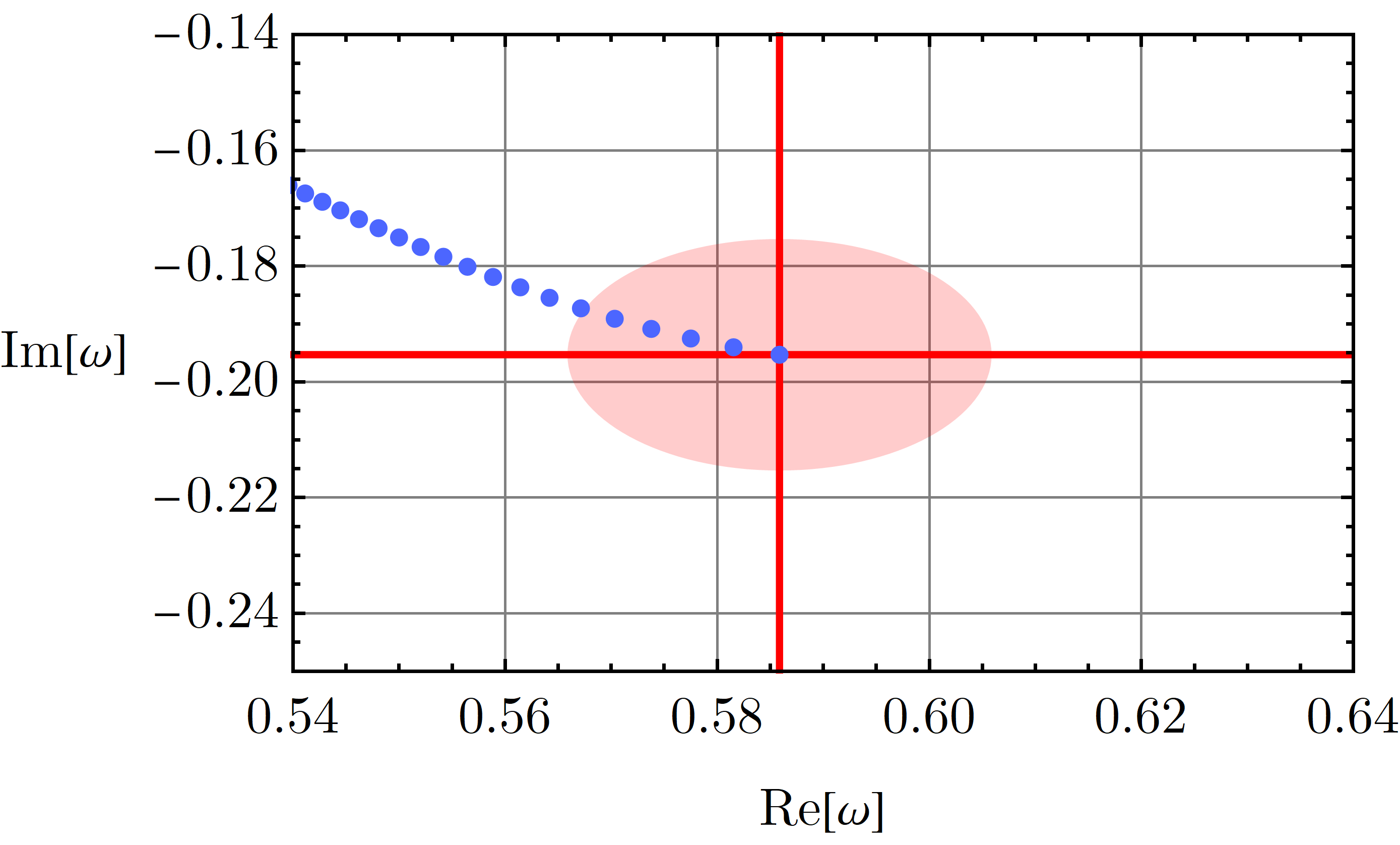}
		\caption{Deviation from GR prediction for the frequency and damping time of (100) as a function of $\alpha_2$ with $\alpha_1=0$. The red shaded region corresponds to a 10\% deviation from the  Schwarzschild value as given in \eqref{con1}. The step size is $\Delta \alpha_2=0.000002$ in all  panels.}\label{l1qnmsa2}
	\end{center}
\end{figure}

Next we compute the $l=1$ fundamental mode for various values of $\alpha_1$ and $\alpha_2$, and again consider a hypothetical 10\% deviation from the Schwarzschild quasinormal frequency $\omega_{100}=0.967287-0.193518i$. The results are shown in \figurename{\ref{l1qnms}} and \figurename{\ref{l1qnmsa2}}.

The scalar $(100)$ mode exhibits interesting behaviour when the leading components of the expansion \eqref{met} deviate from their Schwarzschild values. In particular, when $a_1<1$ the quasinormal mode evolves in a quantitatively similar fashion to the $l=1,m=-1$ mode of the asymptotically flat Kerr black hole as its dimensionless spin parameter $a=J/M^2$ increases away from the Schwarzschild value of $a=0$. This and other modes for scalar perturbations in the Kerr background are shown in \cite{SR:22}.
\\

The above observation suggests that with finite precision, a single measurement of the $(100)$ mode is insufficient to distinguish between a rotating Kerr black hole and a spherically symmetric semi-classical black hole in cases where $a_1<1$. When $\alpha_1>1$ (and  $\alpha_2<-1$) the quasinormal frequency and damping time instead increase relative to the Schwarzschild limit and the mode instead mimics the $l=1,m=1$ Kerr counterpart. Notably, a 10\% deviation in $\omega_{100}$ is achieved by a deviation of just $|\Delta \alpha_1|=0.0001$ relative to the Schwarzschild value, while a comparable shift in $\omega_{100}$ due to spin requires $|\Delta \alpha_1|\approx 0.5$.

\subsection{$l=2$}

\begin{figure}[ht]
	\begin{center}
		\includegraphics[scale=0.55]{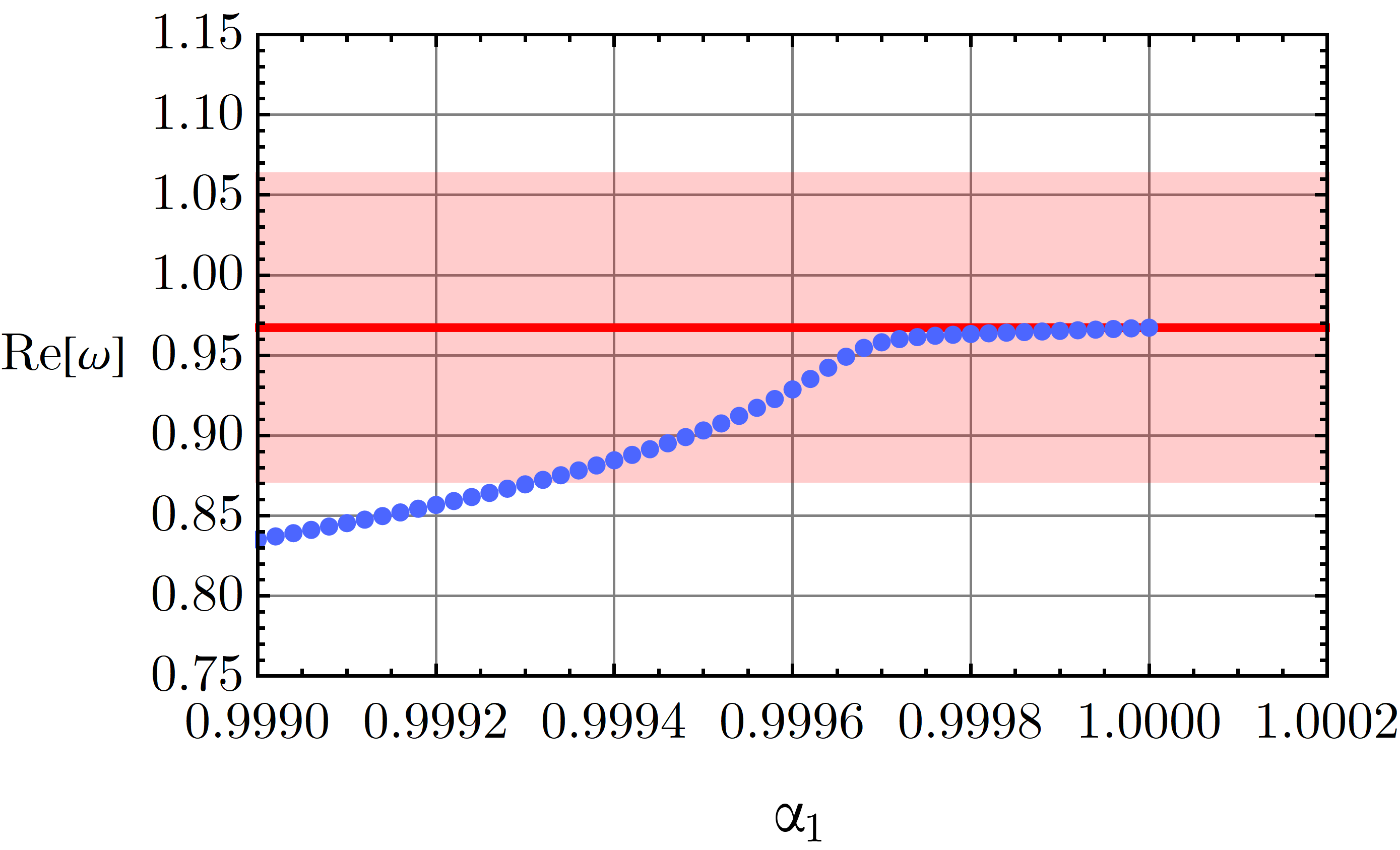}\\	
		\!\!\!\!\includegraphics[scale=0.57]{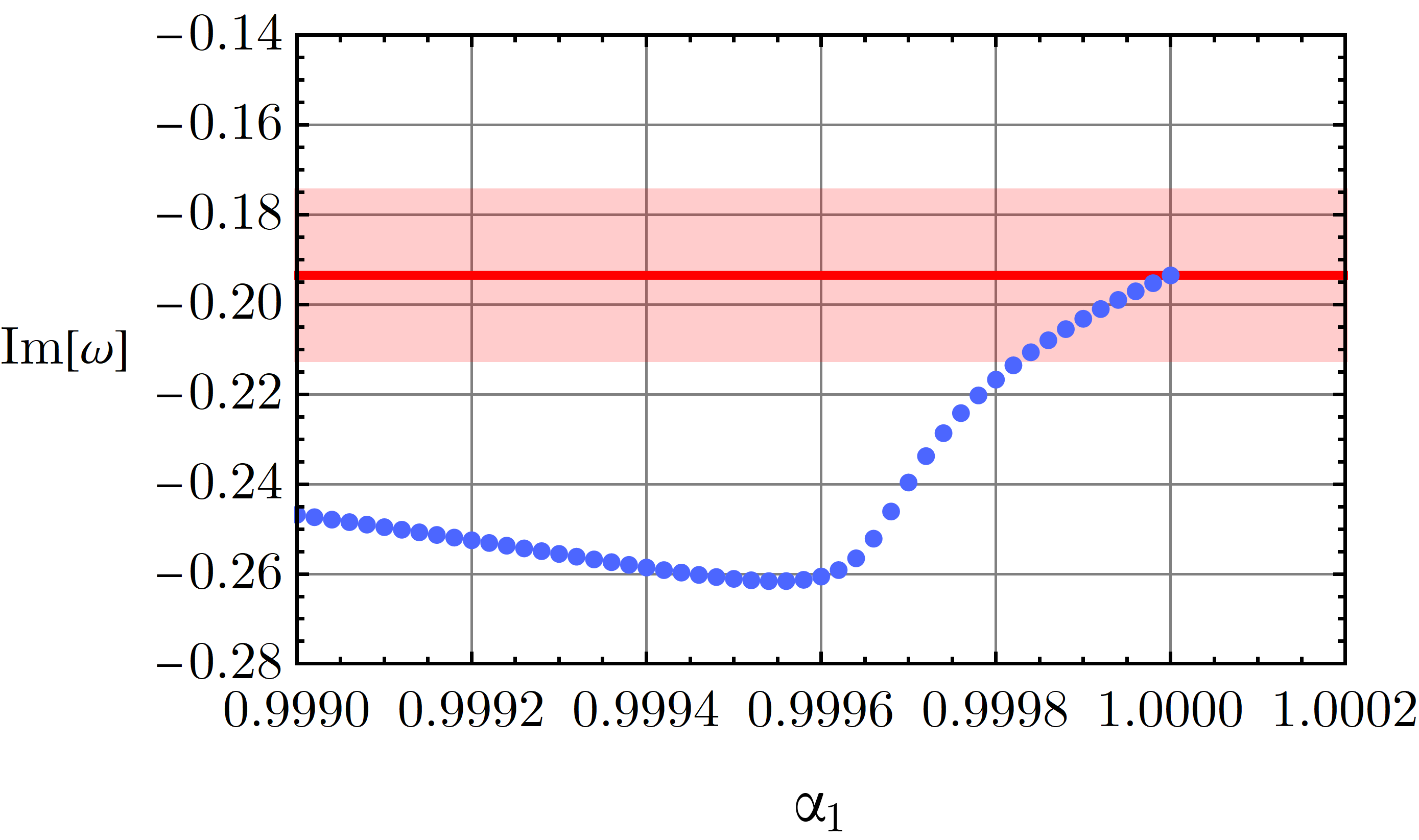}\\
		\!\!\!\!\!\!\includegraphics[scale=0.56]{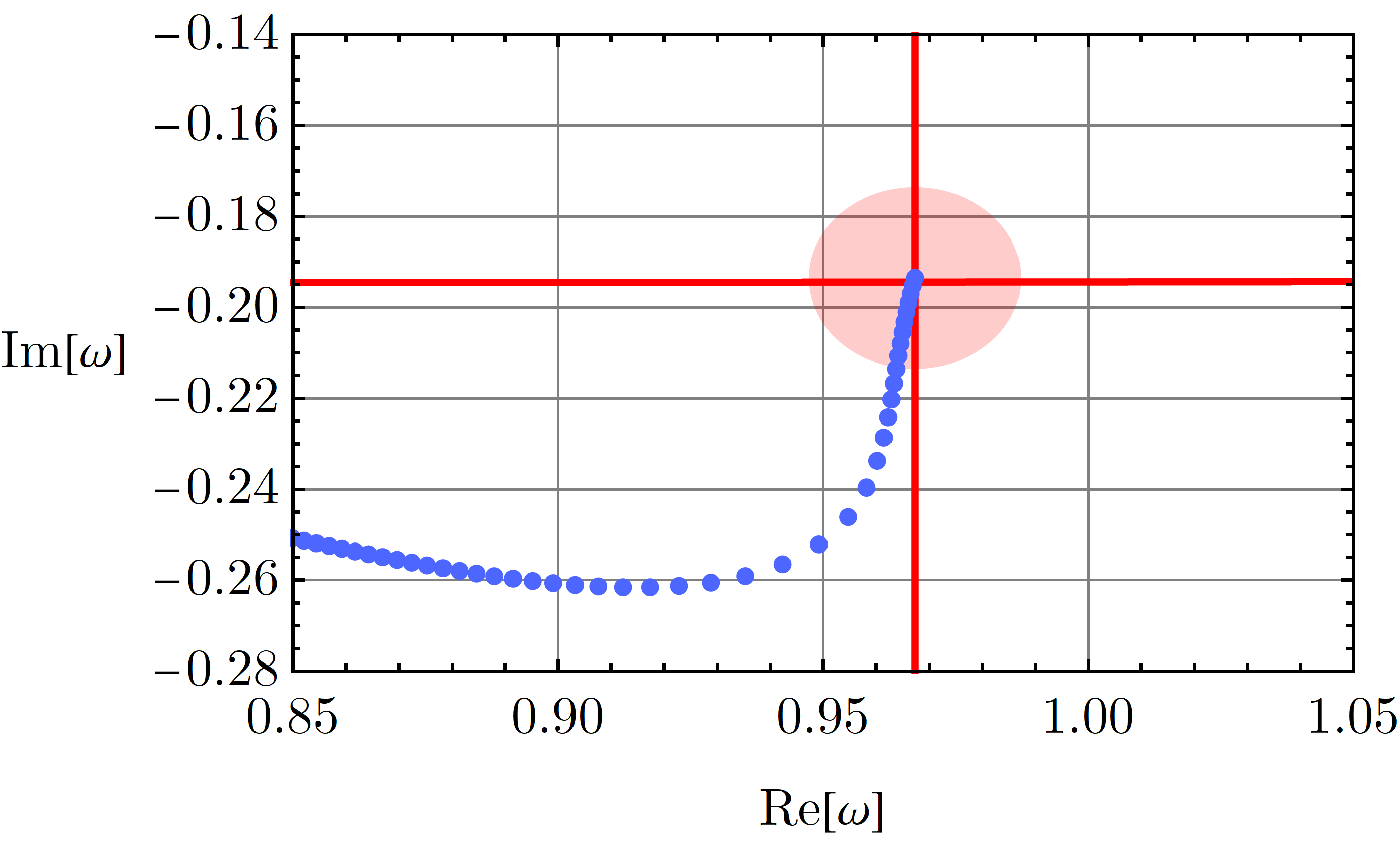}
		\caption{Deviation from GR prediction for the frequency and damping time of (200) as a function of $\alpha_1$ with fixed $a_2=0$. The red shaded region corresponds to a 10\% deviation from the Schwarzschild value. The step size is $\Delta \alpha_1=0.00002$ in all  panels.}\label{l2qnms}
	\end{center}
\end{figure}

\begin{figure}[ht]
	\begin{center}
		\includegraphics[scale=0.55]{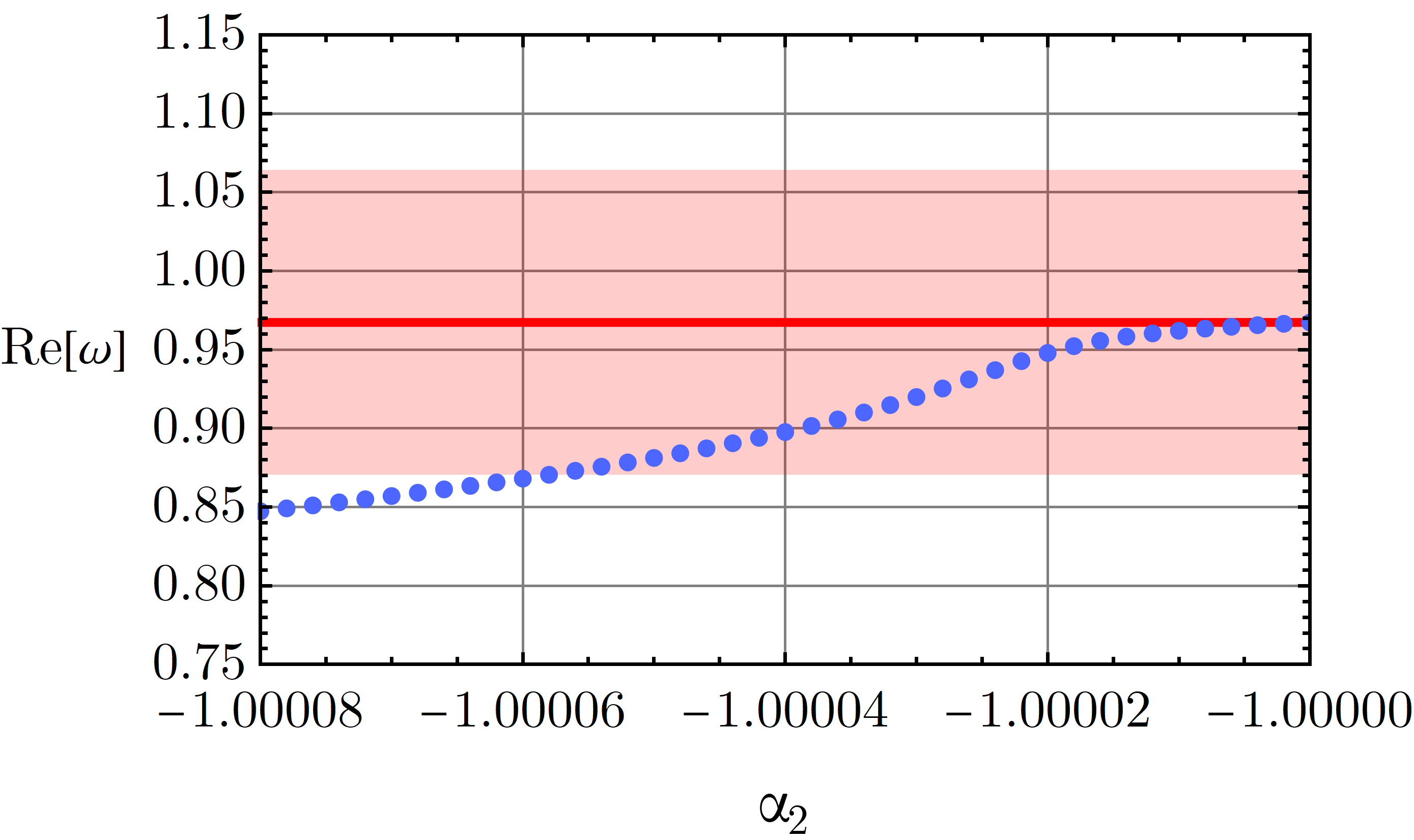}\\	
		\!\!\!\!\includegraphics[scale=0.57]{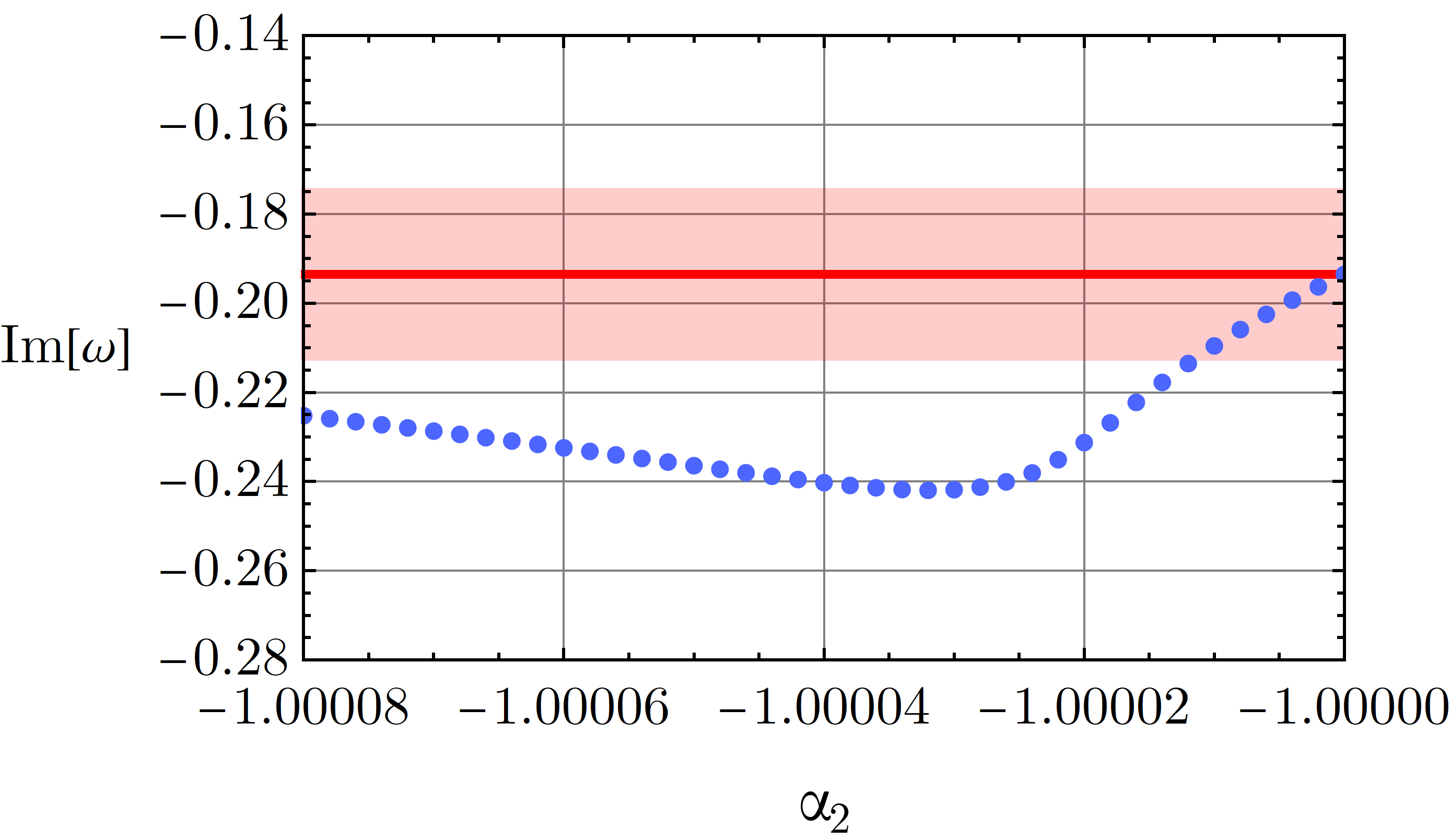}\\
		\!\!\!\!\!\!\!\!\includegraphics[scale=0.54]{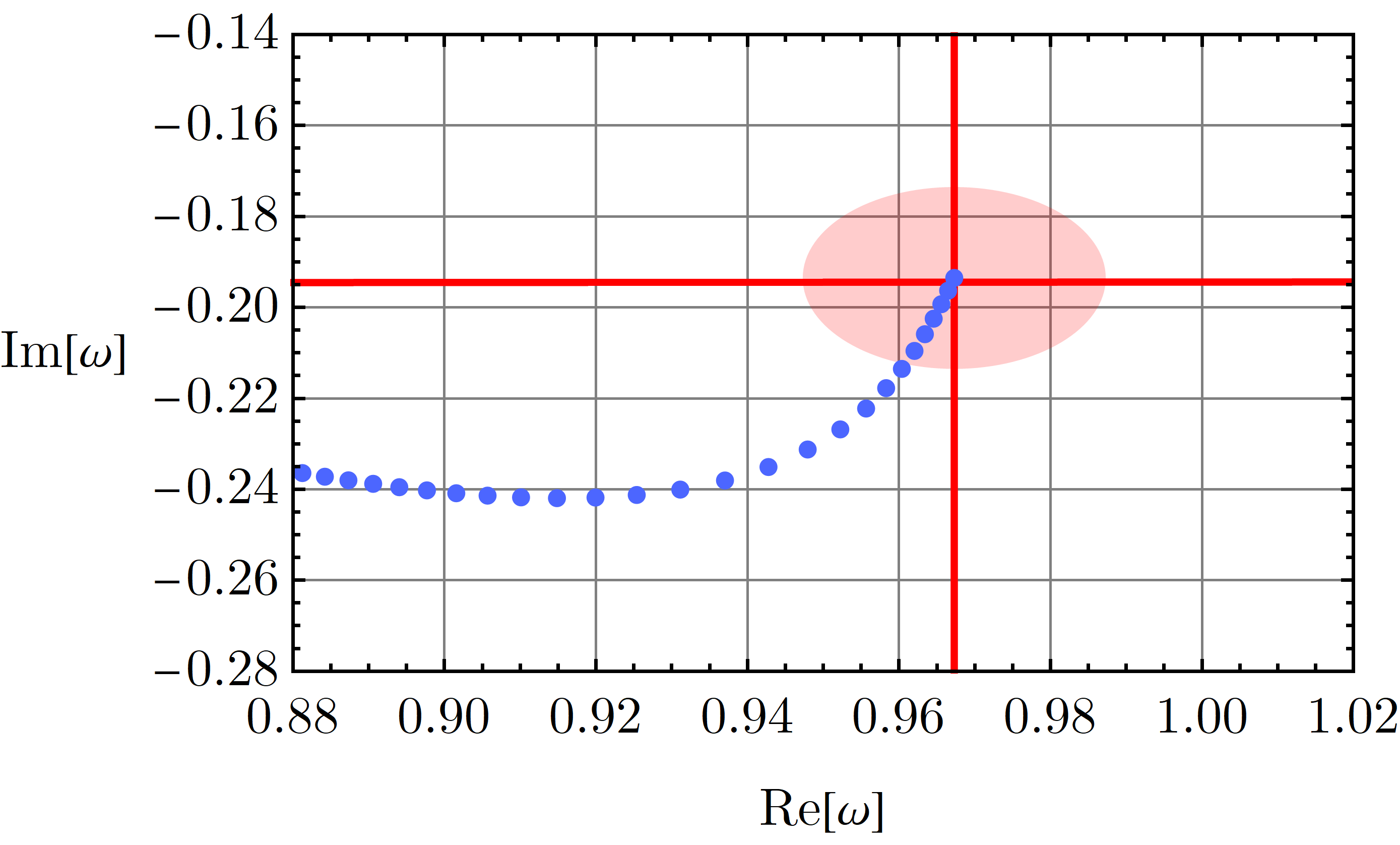}
		\caption{Deviation from GR prediction for the frequency and damping time of (200) as a function of $\alpha_2$ with fixed $\alpha_1=0$. The red shaded region corresponds to a 10\% deviation from the Schwarzschild value. The step size is $\Delta \alpha_2=0.000002$ in all  panels.}\label{l2qnmsa2}
	\end{center}
\end{figure}

We now consider the $l=2$ fundamental mode, which for gravitational perturbations typically dominates the ringdown signal in quasi-circular mergers. We again consider a hypothetical 10\% deviation from the Schwarzschild quasinormal frequency $\omega_{200}=0.967287-0.193518i$. The results are shown in \figurename{\ref{l2qnms}} and \figurename{\ref{l2qnmsa2}}.

All modes display similar sensitivity to underlying changes in $(\alpha_1,\alpha_2)$, with the highest sensitivity to $\alpha_1$ occurring for the frequency of the $(000)$ mode, and the highest sensitivity for $\alpha_2$ being tied to the damping time of $(000)$. A hypothetical maximum deviation of 10\% in their respective quasinormal frequencies translates to a constraint on the leading components of the EMT near the apparent horizon as
\be
E\leq 4.77\times10^{-6}\ ,\qquad e_2\leq 7.16\times 10^{-7}\ .
\ee
These however cannot be expected to vary independently of each other, and further analysis of the sensitivity of the spectrum to the underlying geometry would be required to obtain more precise bounds.

\section{Discussion}

We have computed the quasinormal frequencies of scalar perturbations around a generalized static black hole background. The metric is motivated by the desire to investigate general near-horizon geometries that are constrained only by the requirements of metrics being  static limits of dynamically formed black holes and  sourced by a renormalized energy-momentum tensor whose precise form is constrained by minimal regularity requirements. An interpolation between this semi-classically sourced near-horizon metric and the Schwarzschild metric in the asymptotic region is implemented using a two-point Padé-like approximation, which possesses good analytic properties and has expansions in the near and far regions matching that of the semi-classical and Schwarzschild metrics, respectively.

The low lying modes are computed numerically using a matrix discretization scheme. We consider small deviations from the Schwarzschild geometry near the horizon and compare the resulting quasinormal frequencies to their known Schwarzschild values. The approximation we consider has a smooth limit to the Schwarzschild solution, at which we reproduce the known quasinormal frequencies obtained from a variety of numerical and semi-analytic methods. Small deviations in the leading components of the metric function near the horizon, when taken away from the Schwarzschild limit, lead to a large deviation in the low lying modes form their Schwarzschild values. Using observational constraints on the $l=2$ gravitational mode obtained from LIGO-VIRGO observations as a guide, we assume a hypothetical constraint on the deviation of the scalar modes from their Schwarzschild values, and study the sensitivity of the modes to changes in the underlying geometry.

We find that deviations in the leading coefficient $\alpha_1$ of the order $10^{-4}$ and of the subleading coefficient $\alpha_2$ of order $10^{-5}$ lead to a deviation in the $l\in\{0,1,2\}$ fundamental modes exceeding 10\% of their Schwarzschild values, implying that tight constraints on deviations from the Schwarzschild solution may arise from semi-classical effects near the horizon. Notably, we show that spherically symmetric variations of the near-horizon geometry away from the Schwarzschild metric are able to mimic the imprint that non-zero angular momentum would have on certain scalar quasinormal modes. In principle such an effect has the potential to obscure the distinction between a Kerr black hole and a non-rotating black hole with quantum effects near the horizon in the absence of additional independent measurements, highlighting the importance of multi-modal measurements in distinguishing between various black hole mimickers.

The advantage of our scheme is its relative model-independence. However, one limitation of the two-point Pad\'{e} interpolation lies in the existence of poles in the interpolating function for some (potentially interesting) values of the metric parameters. For example, both functions $f_5(r)$ and $f_4(r)$ cannot accommodate the values of $(\alpha_1,\alpha_2)$ that correspond to the Bardeen metric. This is a drawback that our M-approximants share with the C-approximants used in Ref.~\cite{RZ:14}.  We have shown that in the example we study here, the variation in QNM frequencies from different valid (pole-free) interpolations is appropriately small. We have also considered scalar perturbations in spherical symmetry only, and only in the frequency domain. Finally, while the semi-classical near-horizon metric \eqref{met} describes a wide variety of black hole models in spherical symmetry---such as Reissner-Nordstr\"{o}m and Bardeen regular black holes--the interpolation \eqref{Padé} cannot support these geometries without additional assumptions on the terms entering the asymptotic expansion of the metric.

Future work will  proceed in several directions. First by generalizing to gravitational perturbations, which will be necessary to make concrete comparisons to observational data. We will also aim to develop a more sophisticated interpolation scheme which can incorporate a wider class of metrics, including boundary conditions consistent with the presence of a cosmological constant. Finally we aim to generalize our work to dynamical spacetimes, and compute QNMs for slowly evolving semi-classical black holes in a self-consistent setting. In each case it is desirable to explore as much of the parameter space describing the near-horizon metric as possible, and important classes of problems can only be studied when an interpolating metric can be constructed. Likely a combination of the M-approximants and the C-approximants offers the best solution, and future work aims to develop even more flexible interpolation schemes.

\acknowledgements

F.S. is funded by the ARC Discovery Project Grant No. DP210101279. The work of D.R.T. is supported by the ARC Discovery project Grant No. DP210101279. We thank Roman A. Konoplya for useful comments.

\appendix

\section{The semi-classical metric}\label{appA}

We construct the semi-classical metric and discuss its regularity properties and constrained form (see \cite{MMT:22} for an extensive review). A general spherically symmetric metric in Schwarzschild coordinates (with areal radius $r$) is given by
\be
ds^2=-e^{2h(t,r)}f(t,r)dt^2+f(t,r)^{-1}dr^2+r^2d\Omega_2\ , \label{sgenm}
\ee
while using the advanced null coordinate $v$ results in the form
\be
ds^2=-e^{2h_+(v,r)}f_+(v,r)dv^2+2e^{h_+(v,r)}dvdr+r^2d\Omega_2\ . \label{m:vr}
\ee
The function $f$ is coordinate-independent, i.e. $f(t,r)\equiv f_+\big(v(t,r),r\big)$ and in what follows we omit the subscript. It is conveniently represented via the Misner--Sharp--Hernandez (MSH) mass $M\equiv C/2$ \cite{vF:15} as
\be
f=1-\frac{C(t,r)}{r}=1-\frac{C_+(v,r)}{r}\equiv\partial_{\mu} r \partial^{\mu} r\ ,
\ee
The functions $h$ and $h_+$ play the role of integrating factors in the coordinate transformation
\be
dt=e^{-h}(e^{h_+}dv-f^{-1}dr)\ . \label{trvr-transformation}
\ee
For example, the Schwarzschild metric corresponds to $h\equiv 0$, $C\equiv r_g=\mathrm{const}$, and $v=t+r_*$, where $r_*$ is the tortoise coordinate \cite{FN:98,HE:73}.

It is convenient to introduce the effective EMT components
\begin{align}
	\tau{_t} \equiv &e^{-2h} {T}_{tt}=\tau_t^\mathrm{mat}+\Lambda f/8\pi\ , \label{split1} \\
	\qquad {\tau}{^r} \equiv & T^{rr}=\tau^r_\mathrm{mat}-\Lambda f/8\pi\ , \label{split2} \\
	\tau {_t^r} \equiv & e^{-h}  {T}{_t^r}=\tau_t^{r\mathrm{mat}} \ . \label{split3} 
\end{align}
They allow both for a convenient expression of the regularity conditions and simplify the form of three components of the Einstein equations
\begin{align}
	&\partial_r C = 8 \pi r^2  {\tau}{_t} , \label{eq:Gtt} \\
	&\partial_t C = 8 \pi r^2 e^h  \tau_t^{r}\ , \label{eq:Gtr} \\
	&\partial_r h = 4 \pi r \left(  \tau_t +  \tau^r \right) / f^2\ . \label{eq:Grr}
\end{align}

It can be shown that dynamical black hole solutions belong to one of the classes that are distinguished by the behavior of the EMT components in $(t,r)$ coordinates on approach to the apparent horizon. The $k=0$ solution
corresponds has
\be
\tau_t,\tau^r,\tau_t^r \to -\Upsilon^2,
\ee
as $r\to r_g$ and describes a shrinking black hole (solutions with $\tau_t^r\to +\Upsilon^2$ describe white holes). On the other hand, the EMT components of $k=1$ solutions scale as $\tau_a\propto f^{1}$ as $r\to r_g$. Static solutions, such as Reissner--Nordstr\"{o}m metric, or various models of regular black holes belong to this class. Dynamical regular black hole models, such as Hayward--Frolov or Simpson--Visser models belong to $k=0$ class. In general, almost all spherically-symmetric black hole evolution that occurs in finite time according to a distant observer belongs to the $k=0$ class. It does not include static solutions.

The dynamical black hole solutions of this class in $(t,r)$ coordinates have the near-horizon expansion for $r\gtrsim r_g$
\be\nonumber
C= r_g-4\sqrt{\pi}r_g^{3/2}\Upsilon\sqrt{\tilde x}+\mathcal{O}(\tilde x)\ , \quad h=-\frac{1}{2} \ln{\frac{\tilde x}{\xi}}+\mathcal{O}(\sqrt{\tilde x})\ , \label{k0met}
\ee
where  $\tilde x\defeq r-r_g(t)$, and the function  $\xi(t)$ is determined by the  choice of time variable.	Consistency of the Einstein equations requires that
\be
\frac{d r_g}{dt} {\equiv r'_g} =-\Upsilon\sqrt{\pi r_g\xi}\ .      \label{lumin}
\ee
The functions $\Upsilon(t)$ and $\xi(t)$ have to be found from other considerations, such as matching with the standard results for Hawking radiation. Note that $f\propto \sqrt{x}$ near the horizon.

Our main subjects are the static $k=1$ solutions that we now present. When $\tau_t^r=0$, the leading terms of the effective EMT components are
\be
\tau_t=Ef+\ldots, \qquad \tau^r=Pf+\ldots\ ,
\ee
where we have not yet determined the behaviour of $f$ as a function of the gap $\tilde x$ and have not restricted the allowed (integer or half-integer) powers of $x$ in higher order-terms. At the  horizon, the energy density and pressure are $\rho(r_g)=E$ and $p(r_g)=P$, respectively. From the definition of the outer apparent horizon, we must have that $E\leqslant(8\pi r_g^2)^{-1}$. Having the standard curvature invariants finite at the apparent horizon requires $E=-P$ and the absence of fractional powers of $\tilde x$ at least up to quadratic order.

For $E<(8\pi r_g^2)^{-1}$ the metric functions are
\begin{align}
	&C=r_g+c_1^2 \tilde x+c_2 \tilde x^2+\ldots, \\
	&h=h_1 \tilde x+\ldots,
\end{align}
where $c_1=8\pi r_g^2 E$ (and thus $f=(1-8\pi E r_g^2) \tilde x/r_g +\mathcal{O}(\tilde x^2)$),  the coefficient $h_1=8(e_2+p_2) r_0^2/(1-c_1)^2$ is determined by the coefficients of the  second order terms in the EMT components, etc. If the black hole has no hair whatsoever, then $h\equiv 0$, while for short hair it is a rapidly vanishing function of $\tilde x$.
The next coefficient in the expansion of $f$ is
\begin{align}
\alpha_2&=\frac{1}{3  \left(8 \pi  E\, r_g^2-1\right)^3}\bigg[192 \pi ^2 E\,^2 r_g^4 \left(4 \pi  e_2 r_g^4+3\right)\nonumber\\
	&\quad-24 \pi  E\, r_g^2 \left(8 \pi  e_2 r_g^4+3\right)+12 \pi  e_2 r_g^4\nonumber\\
	&\quad-1536 \pi ^3 E\,^3 r_g^6+3\bigg]\label{c}
\end{align}

\section{Continued fraction representations}\label{appC}

This Appendix compares the M-fraction approximation used in this work to the continued fraction parameterization of Rezzolla and Zhidenko (RZ) \cite{RZ:14}. The two schemes both involve continued fractions but are distinct in a number of important ways. The RZ construction involves parameterizing, or approximating, a known metric by truncating a continued fraction $\tilde{A}(x)$ (Eq. (19a) in \cite{RZ:14}) to a given order. In so doing, one obtains an approximate metric with some pre-determined near-horizon and asymptotic form which can be revealed by expanding the metric near $r=r_g$ and $r=\infty$. The order of approximation in $\tilde{A}(x)$ fixes both of these, and the coefficients $a_i$ defining the continued fraction $\tilde{A}(x)$ appear in both expansions at some finite order. 
\\

By contrast, our scheme involves choosing independently the near-horizon and asymptotic forms of an {\it unknown} metric first, by specifying some finite number of terms in its near and far expansion. The M-fraction then allows one to construct a metric which interpolates between the two expansions, and matches them simultaneously up to the specified order in both the near and far regions. Our construction allows one to specify the form of these expansions {\it independently}, whereas in the RZ scheme, the coefficients $\{a_i\}$ of the continued fraction appear in the coefficients of both the near and asymptotic expansions of the resulting metric. Modifying one of the terms in the near-expansion (for example by changing $a_2$) necessarily also changes the metric in the far region, as $a_2$ appears in those coefficients as well. Thus, the RZ construction does not allow for the independent variation of the near-horizon and asymptotic regions, while the M-fraction construction presented here does. This is desirable for studying deviations of the near-horizon geometry of known black hole solutions without necessarily requiring that such deviations propagate (or can be observed) in the asymptotic region.
\\
\begin{figure}[h]
	\begin{center}
		\includegraphics[scale=0.5]{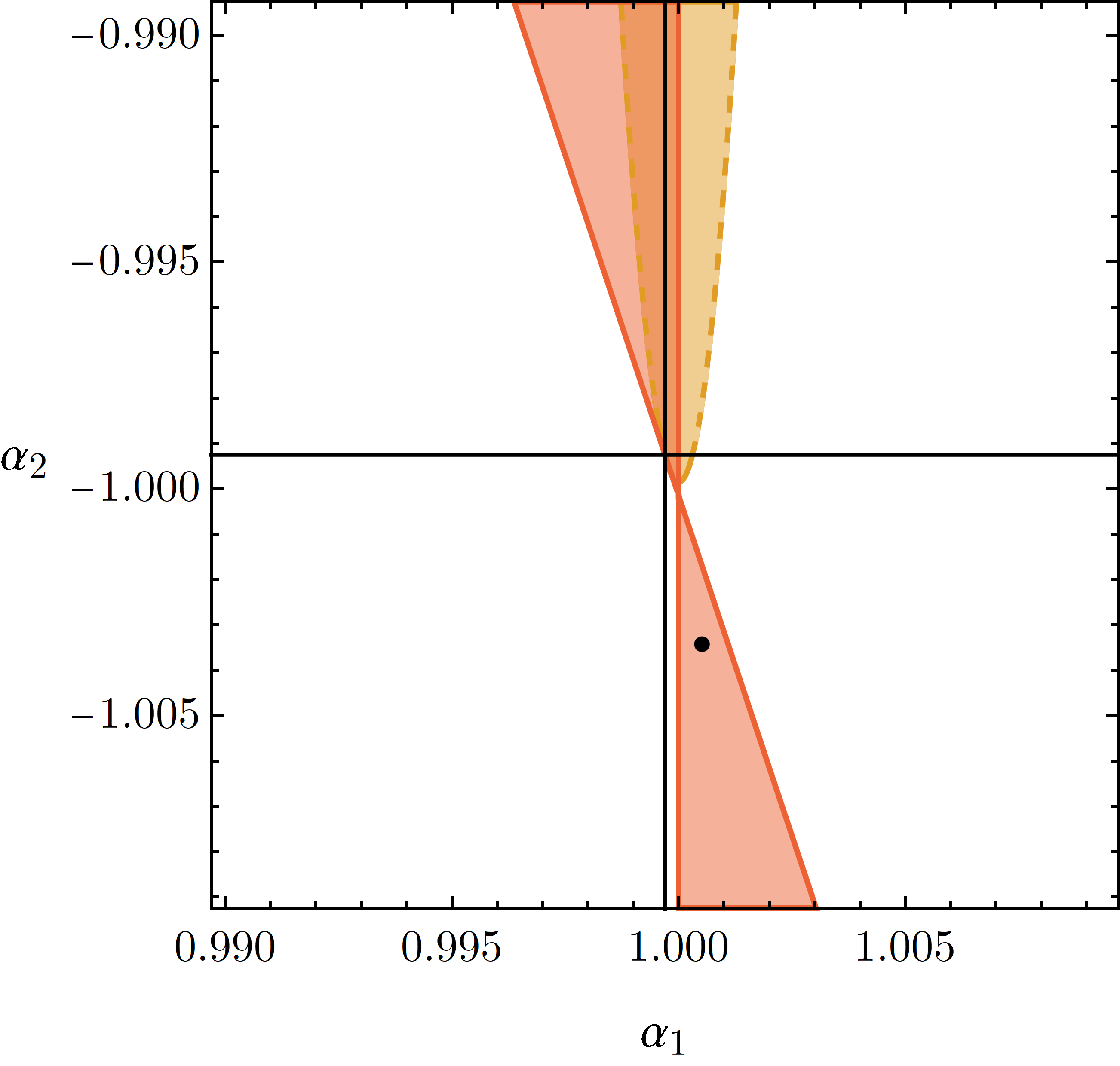}\\	
	\ \includegraphics[scale=0.5]{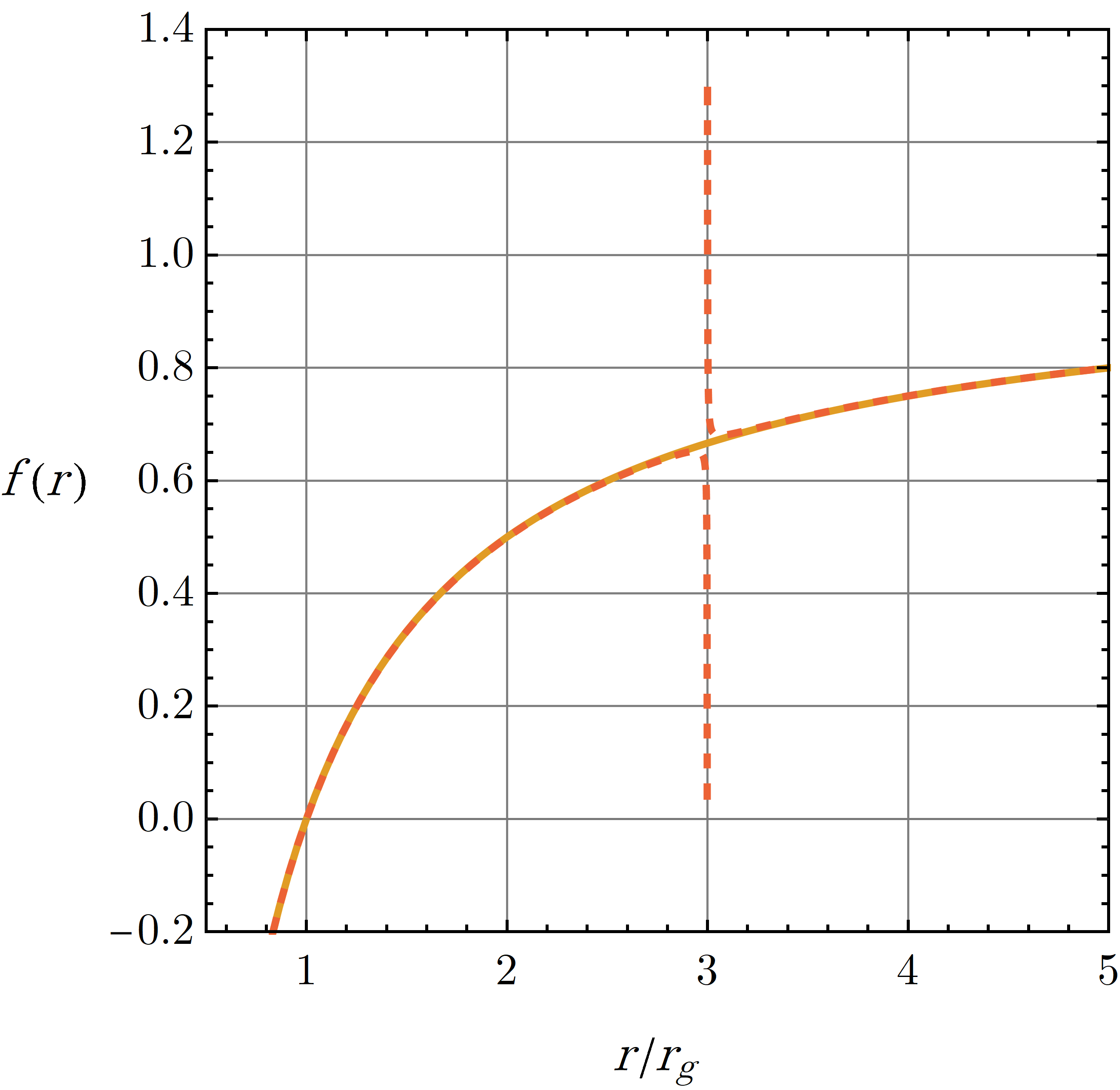}
		\caption{Comparison of the lowest-order M-fraction (yellow) and Rezzolla--Zhidenko (red) approximations of the Bardeen metric \eqref{bd} with $q=0.01$. {\bf Top:} The shaded region corresponds to values of the expansion coefficients ($\alpha_1,\alpha_2$) for which the given approximating metric diverges outside of the horizon. {\bf Bottom:} A plot of the approximating metric function for the metric given by the black point.}\label{compare}
	\end{center}
\end{figure}

In Figure \ref{compare}, we depict the region of validity of the lowest-order non-trivial M-fraction and RZ approximations for the Bardeen metric:
\be\label{bd}
f(r)=1-\frac{r^2 \left(r_g^2+q^2\right)^{3/2}}{r_g^2 \left(r^2+q^2\right)^{3/2}}
\ee
Note the complementary nature of the two approximations. Each point in the figure represents a metric which describes a small deviation away from the Bardeen form, as parameterized by deviations in the leading near-horizon and asymptotic expansion coefficients. For each of the two approximations, the shaded region corresponds to a choice of coefficients which leads to a divergence in the metric function outside of the horizon, rendering the approximation unusable. This is demonstrated by the metric described by the black point in the top panel.
\\

As a simple example, we illustrate the relationship between the parameters of the two schemes. The M-fraction Pad\'{e} expansion of a function with simultaneous expansions \eqref{nearexp} and \eqref{far} is given by adapting Eq.~\eqref{mfrac}. For $n=3$ it is
\begin{widetext}
\be
M_3(\tilde{x})=1-\cfrac{1}{1+\dfrac{1}{r_g}\,\tilde x+\cfrac{\left(\dfrac{\alpha_1- {1}}{r_g}\right) \tilde x}{1-\left(\dfrac{\alpha_1 -1}{r_g}\right)\tilde x+\cfrac{\left(\dfrac{1-\alpha_2 -2 \alpha_1 }{r_g \left(\alpha_1-1\right)}\right) \tilde x}{ 1-\left(\dfrac{1-\alpha_2 -2 \alpha_1 }{r_g \left(\alpha_1-1\right)}\right)\tilde{x}}}}\nonumber
\ee
\end{widetext}
which has a rational function representation as
\be
f_3(x)=\frac{\bar{A}\tilde x^3+\bar{B}\tilde x^2+\bar{C} \tilde x+\bar{D}}{\bar{A}\tilde x^3+\bar{\bar{B}}\tilde x^2+\bar{\bar{C}} \tilde x+\bar{\bar{D}}}
\ee
where the coefficients are related to the continued fraction by
\begin{align*}
\bar{A}&=(1-2 \alpha_1-\alpha_2) \\
\bar{B}&=(\alpha_1 r_g+\alpha_2 r_g) \\
\bar{C}&=\alpha_1 r_g^2 \\ 
\bar{D}&=0\\
\bar{\bar{B}}&=(r_g-\alpha_1 r_g) \\
\bar{\bar{C}}&=r_g^2 \\
\bar{\bar{D}}&= r_g^3\ .
\end{align*}
The two-point Rezzolla--Zhidenko  expansion \cite{RZ:14} of a static spherically-symmetric metric with $h\equiv 0$ which also assumes the validity of GR as we do, is given by
\be\label{rzexp}
f_{\mathrm{RZ}}=(1-2\epsilon)x+3\epsilon x^2-\epsilon x^3+x(1-x)^3\ContFracOp_{m=1}^{n}\left(\frac{{a^C}_m}{1}\right) \ ,
\ee
where the radial variables are related by
\be
\frac{\tilde x}{r_g}=\frac{x}{x-1} \ .
\ee
In terms of the coefficients of Eq.~\eqref{rzexp}, the first two near-horizon expansion coefficients of the metric of Eq.~\eqref{met} are given by
\be
\alpha_1=1+ {a^C_1}-2\epsilon, \qquad \alpha_2=-1-4a^C_1-a^C_1a^C_2+5\epsilon \ .
\ee
If only the two coefficients $a_1^C$ and $a_2^C$ have non-zero values, then $f_{\mathrm{RZ}}$ possesses a pole at
\be
x_2=-1/a_2^C \ ,
\ee
and if the expansion is extended to $n=3$, at
\be
x_3=-1/(a_2^C+a_3^C) \ .
\ee
Requiring that these poles do not occur outside of the horizon determines the domain of validity of the approximation in the parameter space of the metric. For the Bardeen metrics, direct evaluation gives that
\be
a_1^C=-\tfrac{3}{2}\lambda^2+\cO\big(\lambda^4\big) \ , 
\ee
and 
\be
a_2^C=-\frac{3}{2}-\frac{5}{16}\lambda^2+\cO\big(\lambda^4\big)<1 \ ,
\ee
for sufficiently small values of $\lambda$.

\section{The radial wave equation}\label{appB}

The radial wave equation \eqref{radial2} is defined by the coefficients

\begin{figure}[ht]
	\centering
	\hspace{10pt}\includegraphics[width=0.445\textwidth]{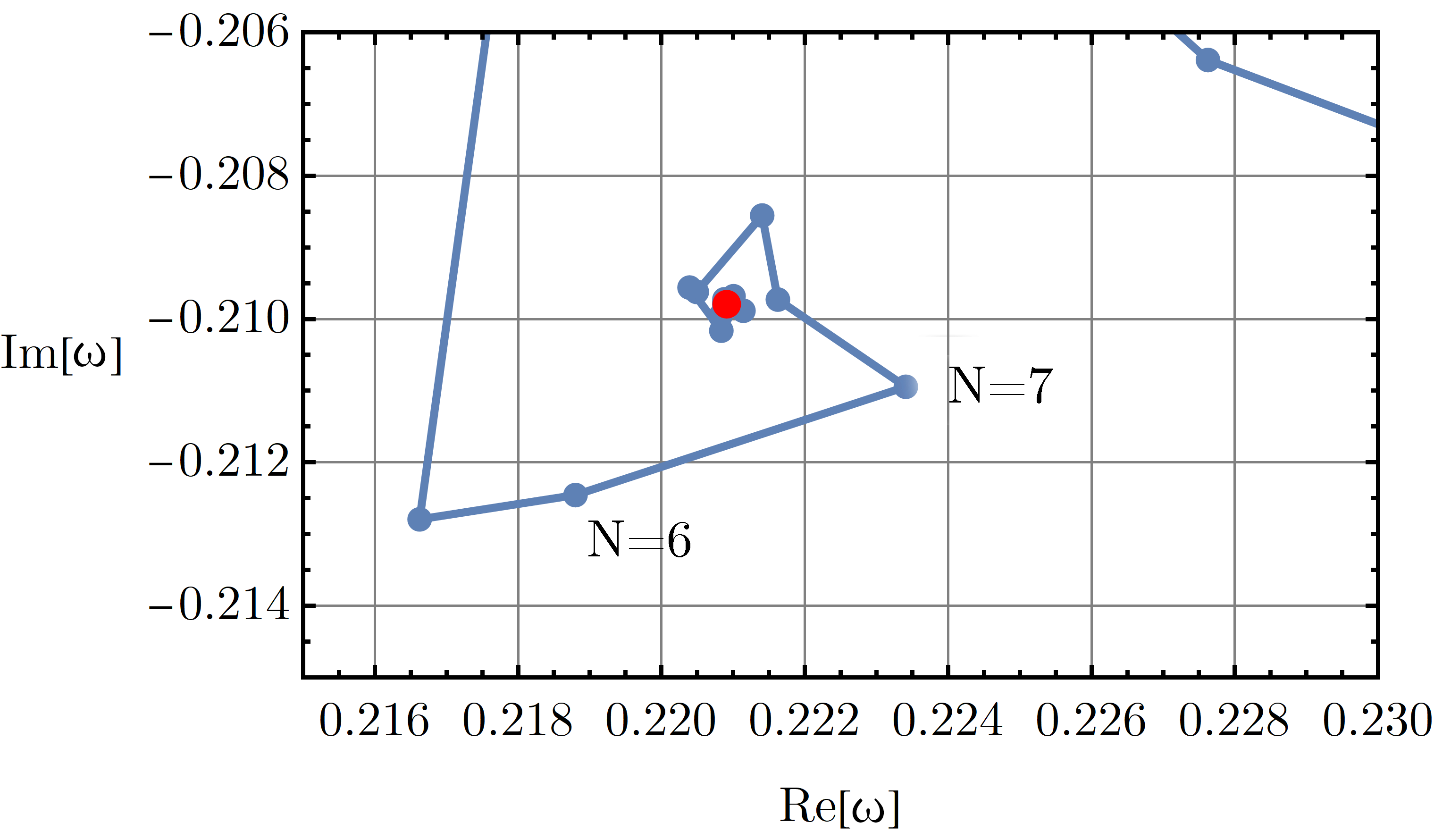}		\includegraphics[width=0.43\textwidth]{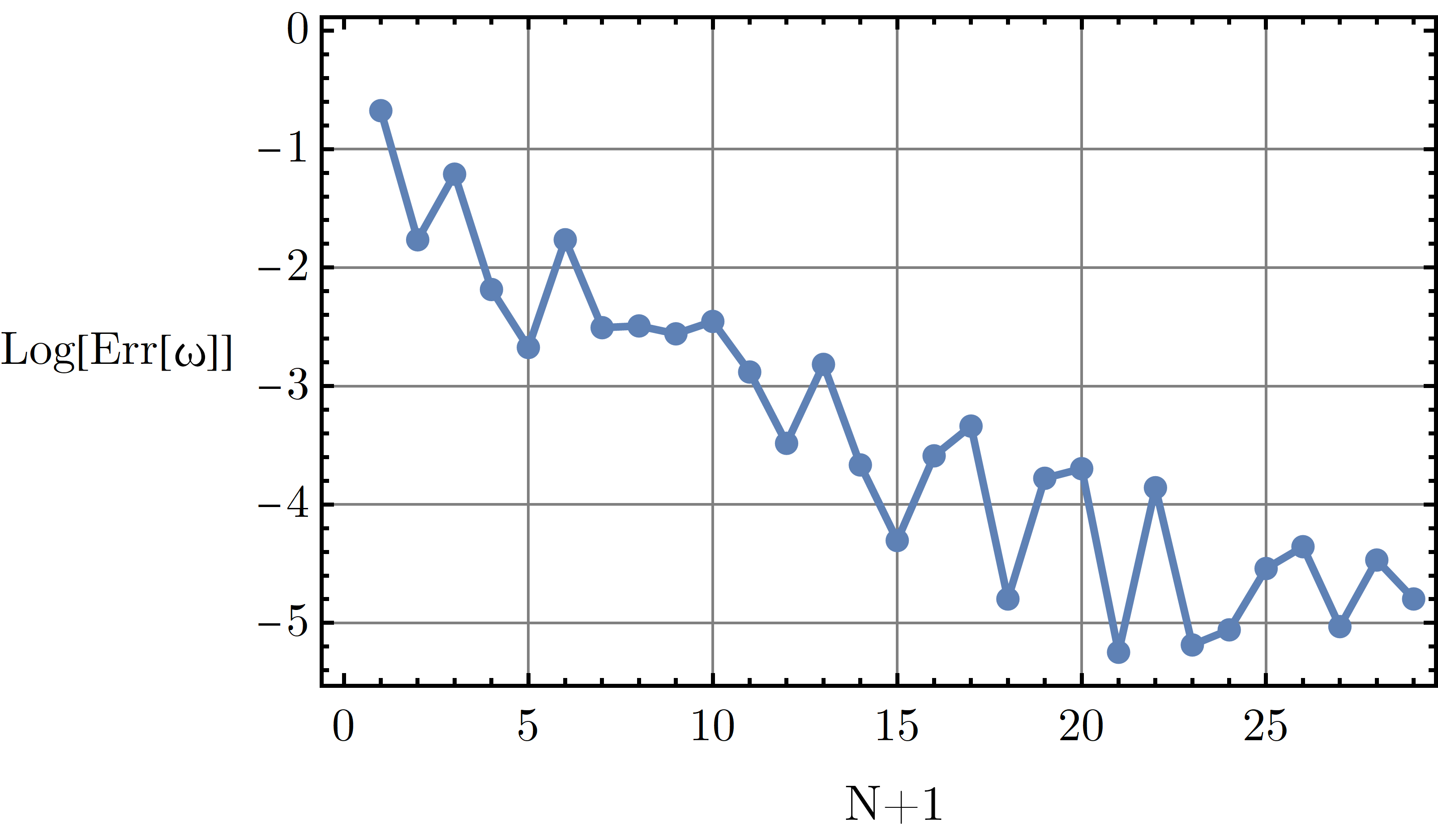}
	\caption{\small{Convergence to known psII and AIM result as reported in \cite{M:22} for the $l=0$ mode, as a function of grid size $N$. The bottom panel shows the logarithm of the relative error between the computed and known result.}}
\end{figure}

\begin{align*}
\sigma(x)&=\frac{1}{(x-1)^9 x^2}\left\{(x-1)^4 (\alpha_1 -1)\left[2 r_g x^5 \omega+  (2 r_g \omega -i)\right.\right.\nonumber\\
&\left.\left.+2 i r_g \omega-x^3 \left(l^2+l-22 r_g^2 \omega ^2+3 i r_g \omega -3\right)\right.\right.\nonumber\\
&\left.\left. +x^2 \left(2 l^2+2 l-16 r_g^2 \omega ^2+5 i r_g \omega -3\right)\right.\right.\nonumber\\
&\left.\left. +x^4 \left(-16 r_g^2 \omega ^2+5 i r_g \omega -1\right)\right.\right.\nonumber\\
&\left.\left.-x \left(l^2+l-10 r_g^2 \omega ^2+7 i r_g \omega -1\right)\right]\right.\nonumber\\
&\left.+x (1-\alpha_1 ) \left(\alpha_2 +1\right)\left[1+4 i r_g \omega\right.\right.\nonumber\\
&\left.\left.+2 x^8 \left(l^2+l+18 i r_g \omega +9\right)+x^9 (-2-4 i r_g \omega )\right.\right.\nonumber\\
&\left.\left.-x \left(2 l^2+2 l-18 r_g^2 \omega ^2+25 i r_g \omega +3\right)\right.\right.\nonumber\\
&\left.\left.-x^7 \left(10 l^2+10 l+18 r_g^2 \omega ^2+117 i r_g \omega +66\right)\right.\right.\nonumber\\
&\left.\left.+x^6 \left(20 l^2+20 l+96 r_g^2 \omega ^2+195 i r_g \omega +132\right)\right.\right.\nonumber\\
&\left.\left.-2 x^3 \left(6 l^2+6 l+27 r_g^2 \omega ^2+25 i r_g \omega +7\right)\right.\right.\nonumber\\
&\left.\left.+x^4 \left(14 l^2+14 l+186 r_g^2 \omega ^2+87 i r_g \omega +81\right)\right.\right.\nonumber\\
&\left.\left.-x^5 \left(20 l^2+20 l+198 r_g^2 \omega ^2+178 i r_g \omega +147\right)\right.\right.\nonumber\\
&\left.\left.+x^2 \left(8 l^2+8 l+2 r_g \omega  (-17 r_g \omega +26 i)\right)\right]\right.\nonumber\\
&\left.-(x-1)^6\left[1+x^3 (-2 r_g \omega +i)^2+2 i r_g \omega \right.\right.\nonumber\\
&\left.\left.+x^2 \left(l^2+l-3 (-2 r_g \omega +i)^2\right)\right.\right.\nonumber\\
&\left.\left.-x \left(l^2+l-8 r_g^2 \omega ^2+8 i r_g \omega +3\right)\right]\right.\nonumber\\
&\left.-(x-1)^3 x \left(\alpha_2 +1\right)\left[1+2 i r_g \omega\right.\right.\nonumber\\
&\left.\left.+x^4 \left(4 r_g^2 \omega ^2-5 i r_g \omega -1\right)\right.\right.\nonumber\\
&\left.\left.-x \left(l^2+l-8 r_g^2 \omega ^2+10 i r_g \omega +2\right)\right.\right.\nonumber\\
&\left.\left.+x^2 \left(2 l^2+2 l+r_g \omega  (-2 r_g \omega +7 i)\right)\right.\right.\nonumber\\
&\left.\left.-x^3 \left(l^2+l+8 r_g^2 \omega ^2-6 i r_g \omega -2\right)\right]\right\}
\end{align*}

\begin{align}
	\tau(x)&=\frac{1}{(x-1)^9}\left\{(x-1)^6 (x+1) \left(-\alpha_2 -1\right) (1+2 i r_g \omega )\right.\nonumber\\
	&\left.-\frac{(x-1)^9+2 i r_g (2 (x-2) x+1) (x-1)^7 \omega }{x}\right.\nonumber\\
	&\left.+\frac{(x-1)^6 (\alpha_1 -1) \left((x-1) x+2 i r_g \left((x-1)^2 x-1\right) \omega \right)}{x}\right.\nonumber\\
	&\left.+(x-1)^3 +(\alpha_1 -1) \left(\alpha_2 +1\right)\left[3 x^2-1\right.\right.\nonumber\\
	&\left.\left.+4 i r_g (x-1)^3 (2 x+1) \omega -2 x^6+12 x^5-24 x^4+22 x^3\right]\right\}
\end{align}

\begin{figure}[ht]
	\centering
	\hspace{10pt}\includegraphics[width=0.42\textwidth]{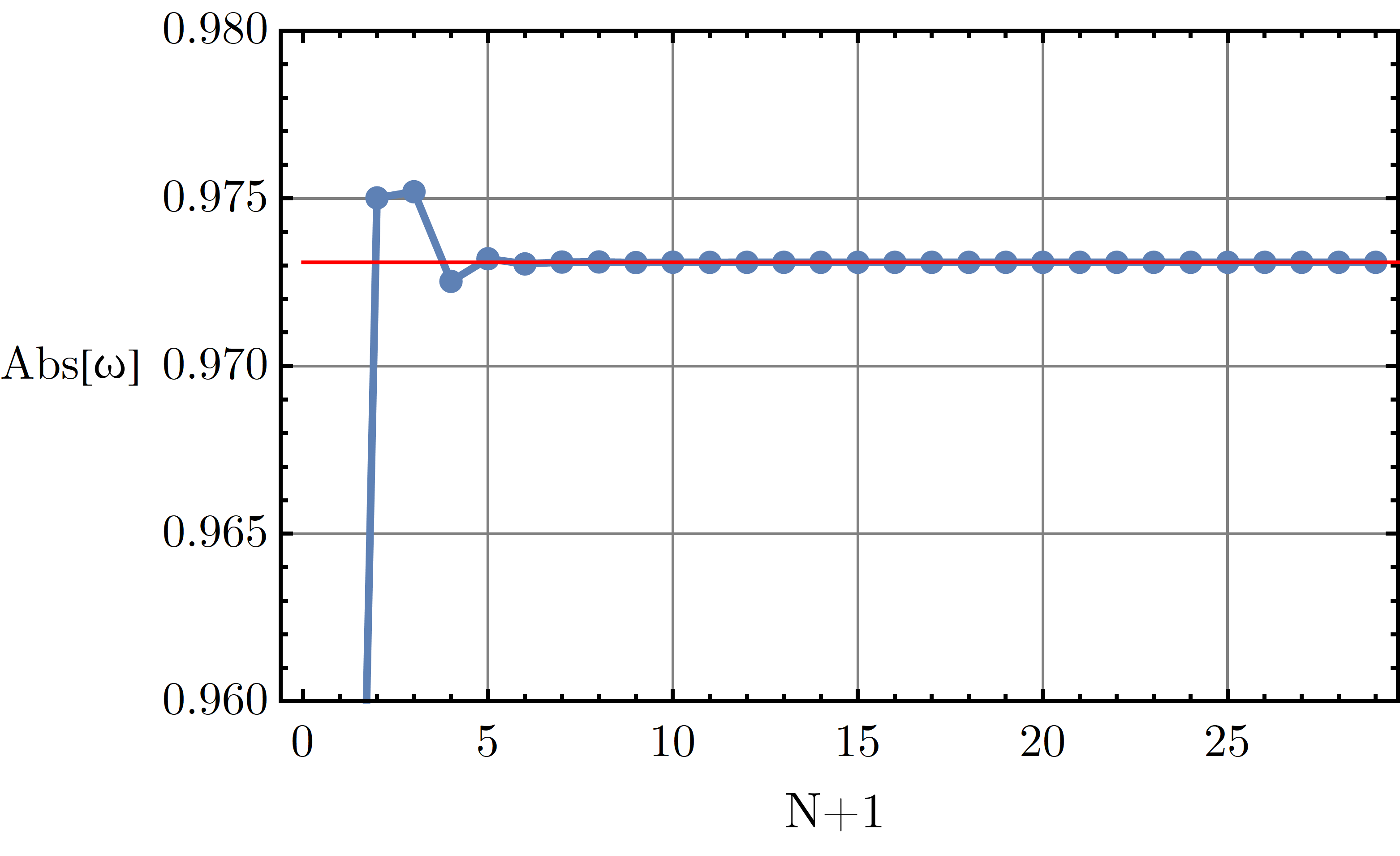}		\includegraphics[width=0.44\textwidth]{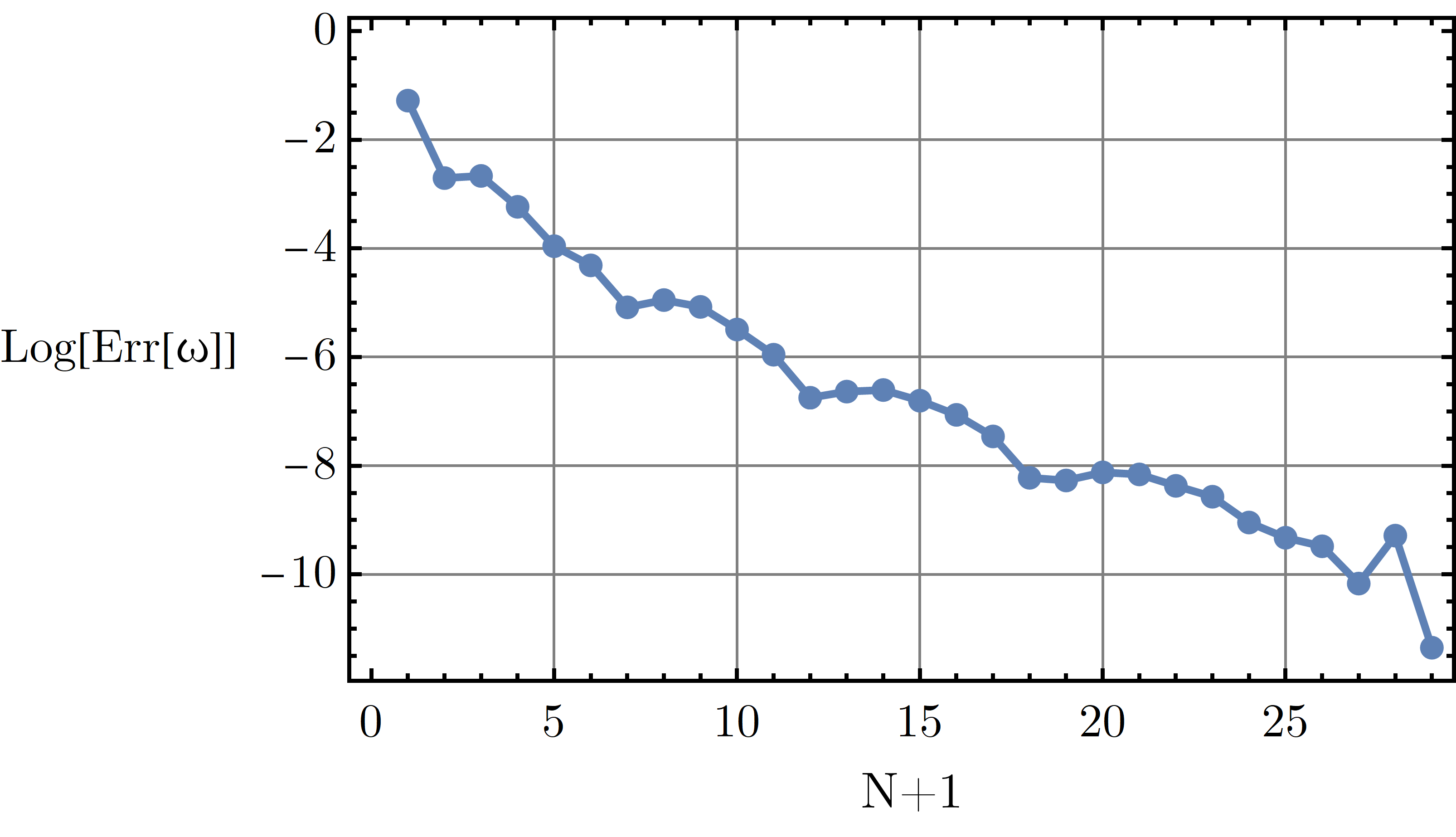}
	\caption{\small{Convergence to known psII and AIM result as reported in \cite{M:22} for the $l=2$ mode, as a function of grid size $N$. The bottom panel shows the logarithm of the relative error between the computed and known result.}}
\end{figure}

\section{Accuracy of matrix method}\label{appD}

We provide some demonstrations of the accuracy and convergence of the matrix method towards known quasinormal frequencies. For example, with a uniformly spaced grid $x_i\in\{0,\tfrac{1}{4},\tfrac{2}{4},\tfrac{3}{4},1\}$, one obtains for the $l=m=0$ fundamental mode of the Schwarzschild black hole:
\be
\omega_{000}=0.21863 - 0.19826 i
\ee
This should be compared to the known frequency obtained through Leaver's or pseudo-spectral methods:
\be
\omega_{000}=0.22091-0.20979i
\ee
With instead a uniformly spaced grid $x_i\in\{0,\tfrac{1}{20},...,\tfrac{19}{20},1\}$, one obtains for the $l=m=0$ fundamental mode
\be
\omega_{000}=0.22088 - 0.20979 i\ .
\ee
Similar accuracy is obtained for higher-$l$ modes. We also compare the absolute and relative errors (in the static case) for the known values of the $l=0$ and $l=2$ modes. These values are obtained from the pseudo-spectral II algorithm with 40 polynomials, as in \cite{M:22}. for the $l=0$ modes, $N>10$ provides a relative error better than $10^{-3}$, while a grid with $N>30$ is required to reach error below $10^{-5}$. The situation is improved for the (more relevant) $l=2$ mode, for which $N=27$ achieves an error as small as $10^{-10}$ relative to the pseudo-spectral II result.


\begin{thebibliography}{100}
	
\bibitem{LIGO:21} LIGO Scientific Collaboration and Virgo Collaboration,
{\href{https://doi.org/10.3847/2041-8213/abe949}{{Astrophys.\ J.\ Lett.} \textbf{913}, L7 (2021)}}.
	

\bibitem{RW:57}T.  Regge and  J. A. Wheeler,
{\href{https://doi.org/10.1103/PhysRev.108.1063}{ Phys. Rev. \textbf{108}, 1063–1069 (1957).}}

\bibitem{P:71} W. H. Press,
{\href{https://doi.org/10.1086/180849}{ ApJ \textbf{170}, L105 (1971).}}

\bibitem{C:92} S.~Chandrasekhar,
	\textit{The Mathematical Theory of Black Holes} (Oxford University Press, Oxford, England, 1992).

\bibitem{FN:98}  V. P. Frolov and I. D. Novikov,
{\href{https://doi.org/10.1007/978-94-011-5139-9}{\textit{Black Holes: Basic Concepts and New Developments} (Kluwer, Dordrecht, 1998).}}

\bibitem{BCNS:19} L.\ Barack, V.\ Cardoso, S.\ Nissanke, and T.\ P.\ Sotiriou (eds.),
{\href{https://doi.org/10.1088/1361-6382/ab0587}{\textit{Class.\ Quantum Gravity} \textbf{36}, 143001 (2019)}}.

\bibitem{RN:03} C. S. Reynolds and M. A. Nowak, 
{\href{https://doi.org/10.1016/S0370-1573(02)00584-7}{{Phys. Rep.} \textbf{377}, 389 (2003).}}
		
\bibitem{B:17} C. Bambi,
{\href{https://doi.org/10.1103/RevModPhys.89.025001}{{ 
Rev. Mod. Phys.} \textbf{89}, 025001 (2017).}}
	
\bibitem{CP:19} V.\ Cardoso and P.\ Pani,
	{\href{https://doi.org/10.1007/s41114-019-0020-4}{ {Living Rev.\ Relativ.} \textbf{22}, 4  (2019)}}.



\bibitem{EHT:19} Event Horizon Telescope Collaboration,
{\href{https://doi.org/10.3847/2041-8213/ab0e85}{{Astrophys.\ J.\ Lett.} \textbf{875}, L4 (2019)}}.
	
\bibitem{IM:19} A. R. Ingram and S. E. Motta, 
{\href{https://doi.org/10.1016/j.newar.2020.101524}{{New Astr. Rev.} \textbf{85}, 101524 (2019).}}
	
\bibitem{HE:73} S. W. Hawking and G. F. R. Ellis,
{\href{https://doi.org/10.1017/CBO9780511524646}{\textit{The Large Scale Structure of  Space-Time} (Cambridge University Press, Cambridge, England, 1973).}}



\bibitem{MMT:22} R. B. Mann, S. Murk, and D.  R. Terno, 
\href{https://doi.org/10.1142/S0218271822300154}{Int. J. Mod. Phys. D \textbf{31}, 2230015 (2022)}.
	


\bibitem{A:20} A.\ Ashtekar,
	{\href{https://doi.org/10.3390/universe6020021}{ {Universe} \textbf{6}, 21 (2020)}}.	
\bibitem{C-R:18} R.\ Carballo-Rubio,
	{\href{https://doi.org/10.1103/PhysRevLett.120.061102}{ {Phys.\ Rev.\ Lett.} \textbf{120}, 061102 (2018)}}.

\bibitem{BBC:21} C. Barcel\'{o} et al.,
{\href{https://doi.org/10.1088/1361-6382/abf89c}{ Class. Quant. Grav. \textbf{38}, 125003 (2021).}}


\bibitem{BMMT:19} V.\ Baccetti, R.\ B.\ Mann, S.\ Murk, and D.\ R.\ Terno,
	{\href{https://doi.org/10.1103/PhysRevD.99.124014}{{Phys.\ Rev.\ D} \textbf{99}, 124014 (2019)}}.

\bibitem{DSST:23} P. Dahal,  F. Simovic, I. Soranidis, and D. R. Terno
{\href{https://doi.org/10.1103/PhysRevD.108.104014}{{Phys.\ Rev.\ D} \textbf{108}, 104014 (2023)}}.



\bibitem{L:20} C. Liu et al.,
{\href{https://doi.org/10.1103/PhysRevD.101.084001}{Phys. Rev. D \textbf{101}, 084001 (2020).}}

\bibitem{G:24} D. M.  Gingrich,
{\href{https://doi.org/10.1103/PhysRevD.109.044044}{ Phys. Rev. D \textbf{109}, 044044 (2024).}}

\bibitem{J:23} S. K. Jha,
{\href{https://doi.org/10.1140/epjc/s10052-023-12123-4}{ Eur. Phys. J. C \textbf{83}, 952 (2023).}}

\bibitem{RZ:14} L. Rezzolla and  A. Zhidenko,
{\href{https://doi.org/10.1103/PhysRevD.90.084009}{ Phys. Rev. D \textbf{90}, 084009 (2014).}}

\bibitem{KZ:22} R. A. Konoplya and A. Zhidenko,
{\href{https://doi.org/10.1103/PhysRevD.105.104032}{Phys. Rev. D \textbf{105}, 104032 (2022).}}

\bibitem{BCS:09} E. Berti,  V. Cardoso, and  A. O. Starinets,
 {\href{https://doi.org/10.1088/0264-9381/26/16/163001}{Class. Quant. Grav. \textbf{26}, 163001 (2009).}}

\bibitem{N:99} H. P. Nollert,
{\href{https://doi.org/10.1088/0264-9381/16/12/201}{ Class. Quant. Grav. \textbf{16}, R159–R216 (1999).}}

\bibitem{KZ:11} R. A. Konoplya and A. Zhidenko,
{\href{https://doi.org/10.1103/RevModPhys.83.793}{ Rev. Mod. Phys. \textbf{83}, 793–836 (2011).}}



\bibitem{BCCD:22} E. Berti et al.,
{\href{https://doi.org/10.1103/PhysRevD.106.084011}{ Phys. Rev. D \textbf{106}, 084011 (2022).}}

\bibitem{FLV:24} E. Franzin, S. Liberati, and V. Vellucci,
{\href{https://doi.org/10.1088/1475-7516/2024/01/020}{ JCAP \textbf{(2024)01}, 020 (2024).}}

\bibitem{K:23} R. A. Konoplya, D. Ovchinnikov, and B. Ahmedov,
{\href{https://doi.org/10.1103/PhysRevD.108.104054}{Phys. Rev. D \textbf{108}, 104054 (2023).}}


\bibitem{B:11}  J. Barranco  et al.,
{\href{https://doi.org/10.1103/PhysRevD.84.083008}{ Phys. Rev. D \textbf{84}, 083008 (2011).}}


\bibitem{H:12} S. Hod,
{\href{https://doi.org/10.1103/PhysRevD.86.104026}{ Phys. Rev. D \textbf{86}, 104026 (2012).}}




\bibitem{MM:76} J. H. McCabe and  J. A. Murphy,
{\href{https://doi.org/http://bura.brunel.ac.uk/handle/2438/1950}{ J. Inst.  Math. Appl. \textbf{17}, 233–247 (1976).}}

\bibitem{S:80} A. Sidi,
{\href{https://doi.org/10.1016/0771-050X(80)90012-1}{ Jour. Comp. App. Math. \textbf{6}, 9–17 (1980).}}

\bibitem{BGM:96} G. A. Baker and  P. Graves-Morris,
{\href{https://doi.org/10.1017/CBO9780511530074}{\textit{Padé Approximants} (Cambridge University Press, Cambridge, England, 1996).}}

\bibitem{CPVWJ:08}   A. Cuyt, V. B. Petersen, B. Verdonk,   H. Waadeland, and  W.  B. Jones,
{\href{https://doi.org/10.1007/978-1-4020-6949-9}{\textit{ Handbook of
Continued Fractions for Special Functions} (Springer, 2008).}}







\bibitem{L:12} T. G. F. Li et al.,
{\href{https://doi.org/10.1103/PhysRevD.85.082003}{ Phys. Rev. D \textbf{85}, 082003 (2012).}}

\bibitem{GVS:12} S. Gossan, J. Veitch, and  B. S. Sathyaprakash.
{\href{https://doi.org/10.1103/PhysRevD.85.124056}{Phys. Rev. D \textbf{85}, 124056 (2012).}}

\bibitem{G:21} A. Ghosh, R. Brito, and  A. Buonanno.
{\href{https://doi.org/10.1103/PhysRevD.103.124041}{ Phys. Rev. D \textbf{103}, 124041 (2021).}}

\bibitem{A:22} R. Abbott et al.,
{\href{https://doi.org/10.48550/arXiv.2112.06861}{Available at arxiv.org/abs/2112.06861 (2021).}}

\bibitem{BT:13} T. Banks and  T. J. Torres,
{\href{https://doi.org/10.48550/arXiv.1307.3689}{Available at arxiv.org/abs/1307.3689 (2013).}}


\bibitem{G:20} S. Gluzman,
{\href{https://doi.org/10.3390/sym12101600}{ Symmetry \textbf{12}, 1600 (2020).}}



\bibitem{LW:02} C. N. Leung and  Y. Y. Y. Wong,
{\href{https://doi.org/10.1119/1.1503379}{ Am. Jour. Phys. \textbf{70}, 1020–1024 (2002).}}




\bibitem{MO:17} J. Matyjasek  and M. Opala,
{\href{https://doi.org/10.1103/PhysRevD.96.024011}{Phys. Rev. D \textbf{96}, 024011 (2017).}}

\bibitem{H:20} M. Hatsuda,
{\href{https://doi.org/10.1103/PhysRevD.101.024008}{Phys. Rev. D \textbf{101}, 024008 (2020).}}

\bibitem{KR:20} P.  Kocherlakota  and L. Rezzolla,
{\href{https://doi.org/10.1103/PhysRevD.102. 064058}{Phys. Rev. D \textbf{102}, 064058 (2020).}}

\bibitem{K:23} R. A. Konoplya,
{\href{https://doi.org/10.1103/PhysRevD.107.064039}{Phys. Rev. D \textbf{107}, 064039 (2023).}}




\bibitem{N:96} H. P. Nollert,
{\href{https://doi.org/10.1103/PhysRevD.53.4397}{Phys. Rev. D \textbf{53}, 4397–4402 (1996).}}

\bibitem{D:20} R. G. Daghigh,  M. D. Green, and J. C. Morey,
{\href{https://doi.org/10.1103/PhysRevD.101.104009}{ Phys. Rev. D \textbf{101}, 104009 (2020).}}

\bibitem{G:72} C. J. Goebel,
{\href{https://doi.org/10.1086/180898}{ ApJ \textbf{172}, L95 (1972).}}

\bibitem{SW:85} B. F. Schutz and C. M. Will,
{\href{https://doi.org/10.1086/184453}{ ApJ. \textbf{291}, L33–L36 (1985).}}
	
\bibitem{IW:87} S. Iyer and C. M. Will,
{\href{https://doi.org/10.1103/PhysRevD.35.3621}{ Phys. Rev. D \textbf{35}, 3621–3631 (1987).}}

\bibitem{J:17} A. Jansen,
{\href{https://doi.org/10.1140/epjp/i2017-11825-9}{ Eur. Phys. J. Plus \textbf{132}, 546 (2017).}}

\bibitem{BV:11} P. Boonserm and  M. Visser,
{\href{https://doi.org/10.1007/JHEP03(2011)073}{ J. High Energ. Phys. \textbf{2011}, 73 (2011).}}

\bibitem{LQ:16} K. Lin and  W.-L. Qian,
{\href{https://doi.org/10.48550/arXiv.1609.05948}{Available at arxiv.org/abs/1609.05948 (2016).}}

\bibitem{LQ:17} K. Lin and W.-L. Qian,
{\href{https://doi.org/10.1088/1361-6382/aa6643}{ Class. Quant. Grav. \textbf{34}, 095004 (2017).}}


\bibitem{LQ:17b} K. Lin, W.-L. Qian, A. B. Pavan, and  E. A. Abdalla,
{\href{https://doi.org/10.1142/S0217732317501346}{ Mod. Phys. Lett. A \textbf{32}, 1750134 (2017).}}

\bibitem{LS:21}  K. Lin, Y.-Y. Sun, and H. Zhang,
{\href{https://doi.org/10.1103/PhysRevD.103.084015}{Phys. Rev. D \textbf{103}, 084015 (2021).}}



\bibitem{SR:22}  P. H. C. Siqueira and  M. Richartz,
{\href{https://doi.org/10.1103/PhysRevD.106.024046}{Phys. Rev. D \textbf{106}, 024046 (2022).}}

\bibitem{M:22} L. A. H. Mamani  et al.,
{\href{https://doi.org/10.1140/epjc/s10052-022-10865-1}{Eur. Phys. J. C \textbf{82}, 897 (2022).}}

\bibitem{vF:15}  V. Faraoni,
{\href{https://doi.org/10.1007/978-3-319-19240-6}{\textit{Cosmological and Black Hole Apparent Horizons}, (Springer, Heidelberg, 2015).}}


	

\end{thebibliography}
\end{document}